\RequirePackage{fix-cm}
\documentclass{svjour3}                     % onecolumn (standard format)
\smartqed  % flush right qed marks, e.g. at end of proof
\usepackage{amsmath,amssymb}
\usepackage{enumitem} %adds item labels e.g. \begin{itemize}[label={--}] \item \end{itemize}
\usepackage{xcolor}

\usepackage{subfigure}

\usepackage{graphicx}

\usepackage{mathptmx}      % use Times fonts if available on your TeX system

\usepackage{multirow}

\usepackage{tabularx}
\newcolumntype{L}[1]{>{\raggedright\arraybackslash}p{#1}}
\newcolumntype{C}[1]{>{\centering\arraybackslash}p{#1}}
\newcolumntype{R}[1]{>{\raggedleft\arraybackslash}p{#1}}
\usepackage{colortbl} % to change color of lines in tables
\usepackage{longtable}
\usepackage{rotating}
\usepackage{lscape}

\usepackage{footnote}
\makesavenoteenv{tabular}
\makesavenoteenv{table}

\usepackage{setspace} % for line spacing (see table 3 for example)

\usepackage{draftwatermark}
\SetWatermarkText{Preprint}
\SetWatermarkLightness{0.9}
\newcommand{\defas}{:=}

\begin{document}

\title{
An all-at-once Newton strategy for methane hydrate reservoir models 
}
% \subtitle{Do you have a subtitle?\\ If so, write it here}


\author{Shubhangi Gupta     \and
        Barbara Wohlmuth    \and 
        Matthias Haeckel    %etc.
}

\authorrunning{S. Gupta \and B. Wohlmuth \and M. Haeckel} % if too long for running head

\institute{S. Gupta \at
              GEOMAR Helmholtz Center for Ocean Research Kiel, Germany \\
              \email{sgupta@geomar.de}           %  \\
           \and
           B. Wohlmuth \at
              Technical University of Munich, Germany
           \and
           M. Haeckel \at
              GEOMAR Helmholtz Center for Ocean Research Kiel, Germany
}


\maketitle

\begin{abstract}
    Marine gas hydrate systems are characterized by highly dynamic transport-reaction processes in an essentially water-saturated porous medium that are coupled to thermodynamic phase transitions between solid gas hydrates, free gas and dissolved methane in the aqueous phase.
    These phase transitions are highly nonlinear and strongly coupled, and cause the mathematical model to rapidly switch the phase states and pose serious convergence issues for the classical Newton's method. 
    One of the common methods of dealing with such phase transitions is the primary variable switching (PVS) method where the choice of the primary variables is adapted locally to the phase state `outside' the Newton loop. In order to ensure that the phase states are determined accurately, the PVS strategy requires an additional iterative loop, which can get quite expensive for highly nonlinear problems.
    For methane hydrate reservoir models, the PVS method shows poor convergence behaviour and often leads to extremely small time step sizes. 
    In order to overcome this issue, we have developed a nonlinear complementary constraints method (NCP) for handling phase transitions, and implemented it within a non-smooth Newton’s linearization scheme using an active-set strategy.
    Here, we present our numerical scheme and show its robustness through field scale  applications based on the highly dynamic geological setting of the Black Sea.
\end{abstract}

\section{Introduction}
\label{sec:introduction}
% \paragraph{}
% % Methane hydrates are crystalline solids formed when water molecules enclathrate methane molecules.
% % Methane hydrates are stable at high pressures and low temperatures. When warmed or depressurised, they decompose into methane and water.
% % In nature, methane hydrates occur in permafrost regions and marine sediments.
% The motivation for research in methane hydrates is multifarious.
% Methane hydrates are seen as an attractive future \textit{energy resource}.
% It is estimated that the total carbon content of methane hydrates is possibly larger than the \textit{combined} carbon content of all other fossil fuels. 
% There are \textit{environmental concerns} that the rising sea temperatures may destabilise marine gas hydrate with serious impacts on global climate and carbon cycle.
% There are concerns regarding the \textit{geotechnical risks} of natural as well as human induced gas hydrate destabilisation.
% % , such as, rapid consolidation and seefloor subsidence, soil fluidisation, local and regional slope instability, etc.

\paragraph{}
The motivation for research in methane hydrates is multifarious.
Methane hydrates constitute a dominant organic carbon pool in the earth system and an important intermediate "capacitor" in the global methane budget. 
Gas hydrates are predominantly formed from biogenic methane that is generated by microbial degradation of organic matter (methanogenesis) in the deep biosphere. 
This methane is migrating upwards as free gas or methane-rich porewater by advection. 
This fluid flow is caused by non-steady state sediment compaction (passive margins), compaction of oceanic sediments during subduction (active margins), and dewatering of minerals at elevated temperatures (passive+active margins).
Over geological times, the hydrates accumulate close to the bottom simulation reflector (BSR, lower stability limit of gas hydrates) because,
the methane flux from below leads to hydrate formation in the gas hydrate stability zone (GHSZ), but the ongoing sedimentation tends to bury the hydrates below the GHSZ where the hydrates dissociate, and the released methane gas migrates back into the GHSZ to re-form the hydrates. 
Towards the seafloor, the hydrates dissolve due to undersaturation of porewaters as a consequence of anaerobic methane oxidation (AOM). 
Some methane gas by-passes the GHSZ and AOM zone if the upward flow is larger than the reaction rates. 
This methane fuels rich cold seep ecosystems.
Our main motivation for modelling the methane hydrate geosystems is to understand this role of gas migration through the GHSZ in the natural carbon cycle.
Methane hydrates are also seen as an attractive future energy resource.
It is estimated that the total carbon content of methane hydrates is possibly larger than the combined carbon content of all other fossil fuels \cite{Pinero2013,Burwicz2011,Archer2009}.
However, there are a number of serious environmental risks associated with the exploitation of gas hydrate reservoirs for the purpose of gas production. 
Our motivation for modelling the methane hydrate geosystems also extends to the feasibility analysis and risk assessment of various gas production scenarios.

\paragraph{}
One of the main challenges in modelling the methane hydrate geosystems comes from the complex phase transitions which cause phases to appear and disappear locally. 
For example, when methane hydrates dissociate due to changes in the local thermodynamic state (i.e., temperature, pressure, and/or salinity conditions), they release methane and water. 
Methane is released as microscopic gas bubbles which, depending on the local solubility limit for methane dissolution, either collapse into the water phase or coalesce leading to the appearance a free gas phase.
Conversely, when methane hydrates form, given the right temperature, pressure, and salinity conditions, they consume methane which may lead to the disappearence of the free gas phase.
The numerical challenges associated with the appearance/disappearance of gaseous or aqueous phases are elaborately discussed in a number of works, like, \cite{Class2006,Marchand2013}.
For methane hydrate models, additional numerical challenges arise.
Firstly, the hydrate and gaseous-aqueous phase transitions are strongly coupled through nonlinear mass and thermal source and sink terms which are highly sensitive to the local thermodynamic state.
Secondly, the hydrate and gaseous-aqueous phase transitions manifest at vastly different time scales.
For the problems on the geological scales, the rate of methane dissolution is many orders of magnitude higher compared to the rate of hydrate phase change. 
Typically, the gaseous-aqueous phase transition is modelled as an equilibrium process, while the hydrate phase transition is modelled as a kinetically driven process.
Together, both these features of the methane hydrate models compound the numerical challenges of the already complex numerics of phase appearance/disappearance in porous media models.

\paragraph{}
A number of different numerical methods have been developed to handle the gaseous-aqueous phase transitions in multi-phase multi-component porous media models, e.g., primary variable switching (PVS) schemes \cite{WuForsyth2001,Class2002}, method of negative saturations \cite{Panfilov2014}, method of persistent variables \cite{Neumann2013,HuangKolditzShao2015}, and non-linear complementary constraints (NCP) approaches \cite{Lauser2011,Krautle2011,GharbiaJaffre2014,BuiElman2018}.
In the most widely used gas hydrate reservoir simulators, e.g., TOUGH-Hydrate \cite{TOUGHHYDRATEv1}, HYRES-C \cite{Janicki2011,Janicki2014}, STOMP-HYD \cite{STOMP}, HRS \cite{HydrateResSimManual_05}, etc., the gaseous-aqueous phase transitions are handled using the PVS schemes, where the choice of the primary variables is adapted locally to the phase state. 
However, due to the strong coupling and nonlinearities, the phase states in gas hydrate models tend to switch back and forth rapidly, and this often leads to spurious oscillation and a drastic reduction in time step size, in the extreme case, even to a breakdown of the numerical algorithm.

\paragraph{}
In this article, we present a robust implicit semi-smooth Newton scheme based on an NCP approach for handling phase transitions in methane hydrate models.
The advantage of an NCP approach is that it ensures that the primary variables of our mathematical model remain the same throughout the simulation, and that the constraints are realized in a variationally consistent manner, resulting in a more robust numerical scheme. 
As a general outline, we cast the inequality constraints arising from the vapour-liquid-equilibrium (VLE) assumption (e.g.,\cite{Helmig1997}) for the $CH_4-H_2O$ system into a set of complementarity conditions which lead to the mathematical structure of a variational inequality (e.g., \cite{Facchinei2013,Tremolieres2011}). 
We reformulate the complementarity conditions as a set of non-differentiable but semi-smooth functions
which are solved together with the governing PDEs of the methane hydrate model fully implicitly using a semi-smooth Newton method (See, e.g., \cite{HagerWohlmuth2010} and the references therein).
We implement our semi-smooth Newton method using an active-set strategy (e.g., \cite{Hintermuller2002,Hueber2005}), where the Jacobian is uniquely determined based on the local phase states which are partitioned into active/inactive sets using the semi-smooth NCP functions.

\paragraph{}
In Sec.\ref{sec:model}, we present our mathematical model and elaborate on the hydrate and the gaseous-aqueous phase transitions.
In Sec.\ref{sec:numerical_solution}, we introduce our numerical solution scheme for handling these phase transitions.
Finally, in Sec.\ref{sec:numerical_examples}, we present some numerical examples to validate our numerical model and to show the robustness of our numerical scheme for realistic field scale applications.
We also make a comparison of the performance of our semi-smooth Newton scheme with that of a PVS scheme. 

% \paragraph{}
% In Sec.\ref{sec:model}, we present our mathematical model and elaborate on the hydrate and the gaseous-aqueous phase transitions.
% In Sec.\ref{sec:numerical_solution}, we introduce our numerical solution scheme for handling these phase transitions.
% Finally, in Sec.\ref{sec:numerical_examples}, we present some numerical examples based on realistic field scale applications.
% First, in Sec.\ref{subsec:example1}, we validate our numerical model by considering a series of phase transitions involving appearance and disappearance of the gas phase, and comparing the solution with that of a PVS scheme. 
% Next, in Sec.\ref{subsec:example2}, we simulate the sedimentation driven gas migration through the GHSZ in the highly dynamic geological setting of the Black Sea, and compare the performance of our semi-smooth Newton scheme against a PVS scheme to show the robustness of our numerical scheme.
% Finally, in Sec.\ref{subsec:example3}, we simulate a gas production scenario, also based on the geological setting of the Black Sea, where we consider that the hydrate phase is randomly distributed within the hydrate layer with saturations ranging between $0-0.6$, and show that our numerical scheme can robustly handle phase transitions even when the phase distributions and the permeability and porosity profiles are highly complex with large variations.

\section{Mathematical Model}
\label{sec:model}
\begin{figure}
 \centering
 \includegraphics[width=\textwidth]{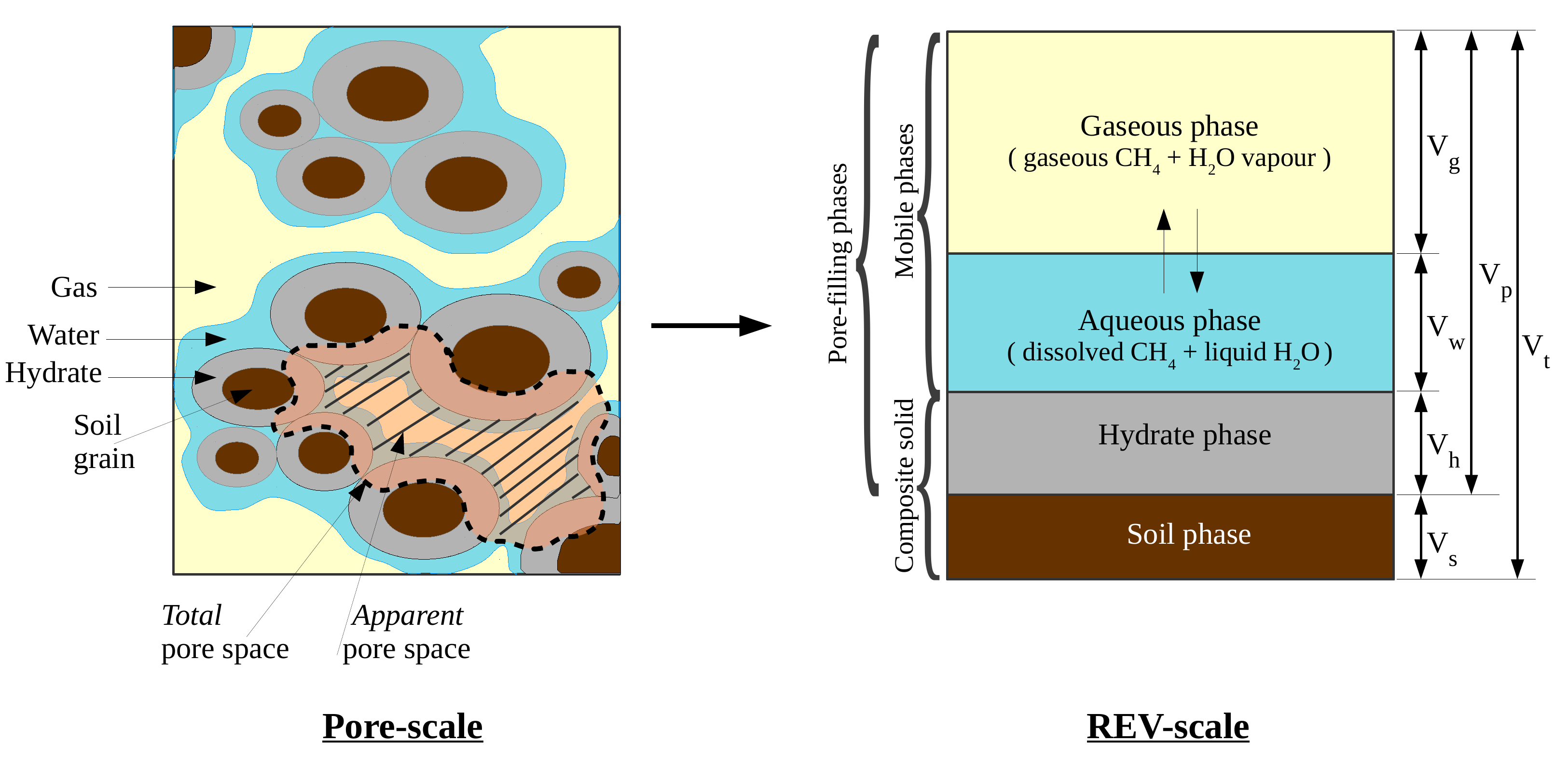}
 \caption{Representation of the phases and components in an REV.
 For any pore-filling phase $\beta=g,w,h$, phase saturation is defined as $S_{\beta}\defas\frac{V_\beta}{V_p}$.
 Total and apparent porosites are defined as $\phi \defas \frac{V_p}{V_t}$, and $\phi_{eff}\defas\frac{V_p-V_h}{V_t}=\left( 1-S_h \right)\phi$, respectively.
  }
  \label{fig:REVconcept}
\end{figure}

\paragraph{}
The model is founded on the theory of porous media and considers the reactive transport processes characterizing a typical methane gas hydrate reservoir on the continuum scale.
The representative elementary volume (REV) underlying the model is shown in Fig. \ref{fig:REVconcept}.

\paragraph{}
The model considers two fluid phases: gaseous and aqueous; and two solid phases: porous granular material (sand or soil) and methane hydrate.
The \textit{phases} are identified with the subscripts $g$, $w$, $s$, and $h$, respectively.
We refer to the sand/soil phase as the \textit{primary sediment matrix} (or simply, the sediment), the hydrate+sediment as the \textit{composite sediment matrix}, and
the fluid and the hydrate phases as the \textit{pore-filling} phases. 
The sediment is assumed to be perfectly rigid. 
The fluid phases are mobile, while the hydrate phase is assumed immobile.

\paragraph{}
The model considers methane hydrate phase change as a kinetic reaction which is strongly dependent on the local thermodynamic state of the system.
The hydrate phase is assumed to contain only pure methane hydrate. Gas ad\-sorption/de\-sorption on the surface of hydrates is not considered.
% The model accounts for methane dis\-so\-lution/exo\-lution, but ignores water evapo\-ration/con\-den\-sation because the methane hydrates typically occur in very high pressure environments.

\paragraph{}
A vast majority of methane hydrate geosystems occur in marine settings where the water salinity has strong influences on the thermodynamics and the phase transitions.
Therefore, the model also considers the transport of dissolved salts. 
% Salt precipitation is, however, not considered.
The model accounts for the miscibility of the fluid phases.
% Note that, the gaseous phase is comprised of only one component: methane; while, the aquesous phase is comprised of three components: methane, water, and salts.
Therefore, the model considers that the gaseous phase is comprised of two components: methane and water; while, the aquesous phase is comprised of three components: methane, water, and salts.
The \textit{components} are identified by the superscripts $CH_4$, $H_2O$, and $c$, for methane, water, and salts, respectively.

\paragraph{}
The model also accounts for the thermal effects, especially the volumetric heat generation due to hydrate phase change, but assumes a local thermal equilibrium within an REV, s.t., a single average temperature can be defined over an REV.

\paragraph{}
In the following, 
subscript '$\alpha$' denotes the fluid phases, while subscript '$\beta$' denotes the pore-filling phases, i.e., $\alpha=g,w$, and $\beta=g,w,h$, and
the superscript '$\kappa$' denotes the components, $\kappa=CH_4,H_2O,c$.
The phase saturations are denoted with $S_{\beta}$, and
mole fractions of each component $\kappa$ in each fluid phase $\alpha$ are denoted with $\chi_\alpha^\kappa$. Note that, $\chi_g^c = 0$.
The fluid phase pressures are denoted with $P_{\alpha}$,
the temperature is denoted with $T$, and
the porosity is denoted with $\phi$
\footnote{$\phi$ refers to the \textit{total} porosity, which indicates the void spaces within the primary sediment matrix.
This is different from \textit{apparent} porosity $\phi_{eff}$, which indicates the actual void spaces available for the fluid flow.
See Fig.\ref{fig:REVconcept} for more details.}.

\subsection{Governing equations}
\label{subsec:governing_equations}

\subsubsection{Mass, momentum and energy conservation}
\label{subsubsec:conservation_laws}

\paragraph{}
The transport processes characterizing the gas production from a typical sub-surface methane hydrate reservoir can be described by invoking the conservation laws for mass, momentum, and energy described for the macroscale properties of the porous medium derived using local volume averaging principles \cite{HassanizadehGray1979a,HassanizadehGray1979b,HassanizadehGray1979c}.

\paragraph{}
Mass balance is considered \textit{component-wise} for each $\kappa=CH_4,H_2O$,
\begin{align}\label{eqn:massBalance_CH4_H2O}
       &        \sum\limits_{\alpha} \partial_t \left( \phi \rho_{\alpha}  \chi_{\alpha}^{\kappa} S_{\alpha} \right)
       \ + \ 	\sum\limits_{\alpha} \nabla\cdot\left( \rho_{\alpha}  \chi_{\alpha}^{\kappa} {\mathbf v_{\alpha}} \right) 
        =	\sum\limits_{\alpha} \nabla\cdot\left(\phi S_{\alpha} {\bf J}_{\alpha}^{\kappa}\right) 
       \ + \ 	\dot g^{\kappa} \ ,
\end{align}
where, ${\mathbf v_{\alpha}}$ denotes the velocity of the fluid phase $\alpha$ relative to the primary sediment matrix, and
${\bf J}_{\alpha}^{\kappa}$ denotes the diffusive flux of the component $\kappa$ in the phase $\alpha$.

\paragraph{}
Mass balance for the hydrate phase is given by,
\begin{align}\label{eqn:massBalance_hydrate}
       \partial_t \left(\phi \rho_{h} S_{h} \right)
 \ = \ \dot g_{h} \ .
\end{align}
In Eqn. \eqref{eqn:massBalance_CH4_H2O} and \eqref{eqn:massBalance_hydrate},
the terms $\dot g^{CH_4}$, $\dot g^{H_2O}$, and $\dot g_{h}$ denote the volumetric source terms resulting from the hydrate phase change.
% Following condition holds: $\dot g^{CH_4} + \dot g^{H_2O} + \dot g^h = 0$ . 

\paragraph{}
Mass balance for the dissolved salt is given by,
\begin{align}\label{eqn:massBalance_salt}
 \partial_t \left(\phi \rho_{w} S_w \chi_w^c \right)
 \ + \ \nabla\cdot\left( \rho_w \chi_w^c {\mathbf v_{w}} \right)
 \ = \ \nabla\cdot\left(\phi S_{w} {\bf J}_{w}^{c}\right) \ ,
\end{align}
where,
${\bf J}_{w}^{c}$ is the Fickian diffusion flux of salt in the aqueous phase.

\paragraph{}
Momentum balance for the fluid phases can be reduced to Darcy’s Law under certain simplifying assumptions (e.g., \cite{Helmig1997}),
\begin{align}\label{eqn:darcysLaw}
  {\bf v}_{\alpha}=-K \ \frac{k_{r\alpha}}{\mu_{\alpha}}\left(\nabla P_\alpha - \rho_{\alpha} {\bf g}\right)  \ ,
\end{align}
where, $K$ denotes the intrinsic permeability of the \textit{composite} sediment matrix, i.e., hydrate+sediment matrix,
$k_{r\alpha}$ denotes the relative permeabilities, and $\mu_{\alpha}$ the dynamic viscosities.

\paragraph{}
The primary sediment matrix is assumed rigid and the hydrate phase is assumed immobile.
Therefore, the momentum of the solid phases is always conserved.

\paragraph{}
For describing the energy conservation in the porous medium, one energy balance equation is sufficient since local thermal equilibrium has
been assumed (e.g., \cite{Helmig1997}).
The energy balance is given by,
\begin{align}
\label{eqn:energyBalance}
    & \partial_t \left[ \left( 1-\phi \right) \rho_s u_s + \sum_{\beta} \left( \phi S_{\beta} \rho_{\beta} u_{\beta} \right)  \right] 
      + \ \sum_{\alpha} \nabla \cdot \left( \rho_{\alpha} {\mathbf v_{\alpha}} h_{\alpha} \right)
    \ = \ \nabla \cdot k^{th}_{eff} \nabla T 
    \ + \ \dot Q_h \ ,
\end{align}
where, $\dot Q_h$ denotes the heat of hydrate phase change.
$h_\alpha$ is the specific enthalpy of fluid phase $\alpha$, $u_\gamma$ is the specific internal energy of any phase $\gamma=g,w,h,s$,
and $k^{th}_{eff}$ is the effective (or lumped) thermal conductivity,
\begin{align*}
  & h_{\alpha} = \int_{T_{ref}}^T Cp_{\alpha}\ dT \ ,
    \notag\\
  & u_{\gamma}  = \int_{T_{ref}}^T Cv_{\gamma }\ dT \ , 
    \notag\\
  \text{and,} \quad
  & k^{th}_{eff} = \left( 1 - \phi \right) k^{th}_s + \sum_{\beta} \left( \phi S_{\beta} k_{\beta}^{th} \right) \ .
\end{align*}

\subsubsection{Closure relationships}
\label{subsubsec:closure_relations}

The phase saturations and the phase pressures are not independent.
The saturations of the pore-filling phases are related through the summation condition,
\begin{align}
\label{eqn:summationcondition_phases}
 \sum_{\beta} S_{\beta} = 1 \ .
\end{align}
The pressures of the fluid phases are related through a capillary pressure $P_c$ as,
\begin{align}
\label{eqn:phasepressuredifference}
 P_g - P_w = P_c(S_{w},S_{h},\phi) \ .
\end{align}
This pressure difference occurs across the gaseous and aqueous phase interface due to balancing of cohesive forces within the liquid and the adhesive forces between the liquid and soil-matrix. 
The parametrization used for approximating $P_c$ is discussed in Sec. \ref{subsubsec:hydraulic_propoerties}.

\subsection{Constitutive relations}
\label{subsec:constitutive_relations}

\paragraph{}
The $9$ governing equations \eqref{eqn:massBalance_CH4_H2O}-\eqref{eqn:phasepressuredifference} consist of the following $25$ unknowns,
$$
S_\beta \ , \
\chi_\alpha^\kappa \ , \
P_\alpha \ , \
P_c \ , \
T \ , \
\mathbf{v}_{\alpha} \ , \
\mathbf{J}_{\alpha}^{\kappa} \ , \
\dot{ g }^{CH_4} \ , \
\dot{ g }^{H_2O} \ , \
\dot{ g }_h \ , \
\dot{ Q }_h \ .
$$
To close the model, we define $16$ additional constitutive relationships in this section for the unknowns
$\chi_\alpha^\kappa$, 
$\mathbf{J}_{\alpha}^{\kappa}$,  
$P_c$, 
$\dot{ g }^{CH_4}$, 
$\dot{ g }^{H_2O}$, 
$\dot{ g }_h$, 
and 
$\dot{ Q }_h$.
Some other properties which are important for modelling hydrate geosystems are also discussed.

\subsubsection{Vapor-liquid equilibrium}

\paragraph{}
Methane and water components are assumed to exist in a state of vapour-liquid equilibrium (VLE), 
and the Henry's law and the Raoult's law are assumed to be valid,
\begin{align}
    \text{Henry's law,}	\qquad z^{CH_4}\chi_{g}^{CH_4} \ P_g &= H^{CH_4}_w \chi_{w}^{CH_4}  \label{eqn:HenrysLaw} \\
    \text{Raoult's law,}\qquad\qquad \chi_{g}^{H_2O} \ P_g &= P^{H_2O}_{sat} \chi_{w}^{H_2O}	 \label{eqn:RaoultsLaw}
\end{align}
where, $z^{CH_4}$ is the methane gas compressibility, $H^{CH_4}_w$ is the pressure-corrected Henry's law solubility constant for methane dissolution in water,
and $P^{H_2O}_{sat}$ is the saturation vapour pressure for water in contact with methane gas.

\paragraph{}
In addition to relationships \eqref{eqn:HenrysLaw} and \eqref{eqn:RaoultsLaw}, we observe that within each phase $\alpha$, the sum of the constituent mole fractions is bounded from above by one, and the equality holds only if the phase is present,
\begin{align}
 \label{eqn:summationConditionForComponents}
 \sum\limits_{\kappa} \chi^{\kappa}_{\alpha} \leq 1 \quad \forall \alpha	
 \qquad \text{and} \qquad
 \sum\limits_{\kappa} \chi^{\kappa}_{\alpha} = 1 \ \text{ iff } \ S_\alpha > 0 \ .
\end{align}
We can cast the conditions in \eqref{eqn:summationConditionForComponents} as a set of Kharush-Kuhn-Tucker complementarity conditions \cite{KuhnTucker1951} as,
\begin{align}
 \label{eqn:complementaryConditions}
 1 - \sum\limits_{\kappa} \chi^{\kappa}_{\alpha} \geq 1 , \qquad S_\alpha \geq 0 , \qquad 	S_\alpha \left( 1 - \sum\limits_{\kappa} \chi^{\kappa}_{\alpha} \right) = 0 , \quad \forall \alpha \ .
\end{align}

\subsubsection{Diffusive mass flux}

\paragraph{}
The diffusive solute flux through the composite sediment matrix is evaluated using Fick’s Law (e.g., \cite{Helmig1997}), 
\begin{align}\label{eqn:FicksLaw}
  & {\bf J}_{\alpha}^{\kappa} = - \tau D_{\alpha}^{\kappa} \left( \rho_{\alpha} \nabla \chi_{\alpha}^{\kappa} \right) \ ,
\end{align}
where, $\tau$ denotes the tortuosity of the composite sediment matrix, 
and $D_{\alpha}^{\kappa}$ are the molecular diffusion coefficients for components $\kappa$ through fluid phases $\alpha$.
Additionally, the summation conditions $\ \sum\limits_{\kappa}{\bf J}_{\alpha}^{\kappa} = 0 \ $ hold for all phases $\alpha$.
Note that, ${\bf J}_{g}^{c} = 0$ since $\chi_g^c = 0$.

\subsubsection{Hydrate phase change kinetics}

\paragraph{}
When solid methane hydrates are warmed or depressurized, they decompose into methane gas and liquid water, and vice versa. 
This chemical reaction is expressed as
$CH_4 \cdot N_{h} H_2O \rightleftharpoons CH_4 + N_h H_2O$, where,
$N_h$ gives the stoichiometry of water molecules per molecule of gas, i.e.  the hydration number. 
The rate of this reaction is modeled by the Kim-Bishnoi kinetic model \cite{KimBishnoi1987},
where, the rate of methane and water generated as a result of hydrate phase change are evaluated as,
\begin{align}\label{eqn:methaneGenerationRate}
 & \dot g^{CH_4} = k^{r} M^{CH4} A_{rs} \left( P_e - P_g \right) 
 \\ \label{eqn:waterGenerationRate}
 & \dot g^{H_2O} = \dot g^{CH_4} N_{h} \frac{M^{H_2O}}{M^{CH_4}} \ ,
\end{align}
where, $P_e$ is the equilibrium pressure for the methane hydrate,
$k^{r}$ is the kinetic rate constant, and
$A_{rs}$ is the specific reaction surface area.
$M^\kappa$ denotes the molar weights, and for methane hydrate, $M_h = M^{CH_4} + N_h M^{H_2O}$. 
Additionally, the following condition holds, 
\begin{align}
\label{eqn:sourceTermsSummationCondition}
 \dot g^{CH_4} + \dot g^{H_2O} + \dot g_h = 0 \ .
\end{align}

\paragraph{}
For the hydrate phase change, the following constraints are considered:
Hydrate dissociation can occour only when hydrate is available and the gas phase pressure is lower than the hydrate equilibrium pressure, and conversely, hydrate formation can occour only when both water and gaseous methane are available and the gas pressure is higher than the hydrate equilibrium pressure,
\begin{align}
 \label{eqn:hydrateDissociationConstraints}
 & \dot g_{h} < 0 \qquad \text{iff} \qquad P_g < P_e \text{ and } S_h > 0 \ , \\
 \label{eqn:hydrateFormationConstraints}
 \text{and,} \quad 
 & \dot g_{h} > 0 \qquad \text{iff} \qquad P_g > P_e \text{ and } S_g>0 \ , \ S_w > 0 \ .
\end{align}
At $P_g=P_e$, no reaction can occour, i.e., $\dot g_{h} = 0$, irrespective of the phase distributions.
Also, from Eqns. \eqref{eqn:methaneGenerationRate}, \eqref{eqn:waterGenerationRate}, and \eqref{eqn:sourceTermsSummationCondition}, it follows that,
$$
\dot g_h > 0	\quad \implies \quad \dot g^{CH_4} < 0 \ , \ \dot g^{H_2O} < 0 \ ,
$$
and vice-versa. 
In the Kim-Bishnoi kinetic model (Eqn.\ref{eqn:methaneGenerationRate}), the constraints \eqref{eqn:hydrateDissociationConstraints} and \eqref{eqn:hydrateFormationConstraints} are ensured through,
\begin{align} \label{eqn:reactionSurfaceArea}
 k^r > 0 \qquad \text{and,} \qquad
 A_{rs} &= \Gamma_r A_s \\
 \text{s.t., }    
    A_{s} &= A_0 \left( 1-S_h \right)^n \quad \text{with } n>0 \ ,  \notag \\
 \text{and, }
     \Gamma_r &= \left\{
      \begin{array}{ll}
       S_h 		& \text{\quad for } \left( P_e - P_g \right) > 0 	\notag \\
       S_g S_w 		& \text{\quad for } \left( P_e - P_g \right) \leq 0  \ ,\notag
     \end{array}
   \right.
\end{align}
where, $A_s$ denotes the specific surface are of the composite sediment matrix, while $A_0$ denotes the specific surface are of the primary sediment matrix. 
Note that, in the limit of $S_h = 1$, i.e., fully clogged pores, no reaction will occur in either direction due to unavailability of reaction surfaces. 
Within the scope of this work, we do not consider this limit.

\paragraph{}
Hydrate dissociation is an endothermic process, and conversely, hydrate formation is an exothermic process.
The heat of reaction associated with the hydrate phase change is commonly modelled as empirical functions of the form (e.g., \cite{Kamath1984}),
\begin{align}
 \label{eqn:hydrateHeatOfReaction}
 \dot Q_h = \frac{\dot g_{h}}{M_h} \left( a_1 + a_2 T\right) \ .
\end{align}

\subsubsection{Hydraulic properties}
\label{subsubsec:hydraulic_propoerties}

\paragraph{}
The capillary pressure $P_c$ of the composite sediment matrix is modelled as \cite{Gupta2015},
\begin{align}
\label{eqn:Pc_scaled}
& P_c = P_{c0}\cdot f^{Pc}_{S_h}\left(S_h\right) \\
\text{where, } 
& P_{c0} = p_{0} S_{we}^{\ -1 / \lambda} \notag \\
\text{and, }
& f^{Pc} = \left(1-S_h\right)^{-\frac{m \lambda - 1}{m \lambda}} \ . \notag
\end{align}
In Eqn.\eqref{eqn:Pc_scaled}, $P_{c0}$ denotes the capillary pressure of the primary sediment matrix,
and $f^{Pc}$ denotes the scaling factor which accounts for the effect of changing effective pore space due to hydrate phase change.
$P_{c0}$ is parameterized using the Brooks-Corey \cite{BrooksCorey1964} model, where
$p_{0}$ is the gas phase entry pressure, 
$\lambda$ is the soil specific parameter depending on the pore-size distribution, and 
$S_{we}$ is the normalized aqueous phase saturation,
$S_{we} = \dfrac{S_w - \left(S_{wr}+S_{gr}\right)}{1 - S_h - \left(S_{wr}+S_{gr}\right)} $,
where, $S_{wr}$ and $S_{gr}$ are the irreducible aqueous and gas phase saturations, respectively.

\paragraph{}
The relative fluid phase permeabilities are also parameterized following the Brooks-Corey model,
\begin{align}\label{eqn:BCRelativePermeabilities}
  k_{rw} = S_{we}^{\frac{2+3\lambda}{\lambda}} 
 \quad \text{ and } \quad 
  k_{rg} = \left( 1-S_{we} \right)^{2} \left( 1 - S_{we}^{\frac{2+\lambda}{\lambda}}\right) \ .
\end{align}

\paragraph{}
The intrinsic permeability of the composite sediment matrix is modelled as,
\begin{align}
\label{eqn:K_scaled}
& K = K_{0}\cdot f^{K}\left(S_h\right) \\
\text{where, }
& f^{K} = \left(1-S_h\right)^{\frac{5m + 4}{2m}} \, \notag
\end{align}
In Eqn.\eqref{eqn:K_scaled}, $K_0$ is the intrinsic permeability of the primary sediment matrix, 
and $f^{K}$ is the scaling factor which accounts for the effect of changing effective pore space due to hydrate phase change.
The scaling factors $f^{Pc}$ and $f^{K}$ were derived \cite{SGuptaThesis2016} based on the assumption that hydrate grows uniformly along the pore surfaces. 
Factor $m$ is a measure of the \textit{sphericity} of the hydrate growth.
In general settings, $0 < m \leq 3$.
For the ideal case of a spherical growth, $m=3$.
The more the hydrate growth skews in the direction of the grain contacts, the lower is the $m$ value. 
For example, according to the experimental investigations by \cite{Kossel2018}, for hydrates formed in quartz sand, $K = K_0 \left(1-S_h\right)^{11.4}$ implying that $m=0.225$.

% \subsection{Phase transitions}
% \label{subsec:phase_transitions}
% 
% \input{model_phase_transitions}

\subsection{Primary variables}
\label{subsec:primary_variables}

\paragraph{}
To solve the mathematical model numerically, we substitute the Darcy velocity (Eqn.\ref{eqn:darcysLaw}) and the constitutive relationships \eqref{eqn:FicksLaw}-\eqref{eqn:hydrateHeatOfReaction} in the coupled  system of Eqns. \eqref{eqn:massBalance_CH4_H2O}-\eqref{eqn:massBalance_salt}, and \eqref{eqn:energyBalance}.
This results in a highly nonlinear system with $11$ unknowns:
$
\left( P_g, P_w , S_g, S_w, S_h , \chi_w^c , \chi_g^{CH_4} , \chi_w^{CH_4} , \chi_g^{H_2O} , \chi_w^{H_2O} , T \right) \ .
$
% $
% \left( P_\alpha , S_\beta , \chi_w^c , \chi_\alpha^{CH_4} , \chi_\alpha^{H_2O} , T \right) \ .
% $

\paragraph{}
Eqn. \eqref{eqn:summationcondition_phases} provides an additional relationship for the phase saturations, reducing the number of unknown saturations to $2$.
Eqn. \eqref{eqn:phasepressuredifference} provides a relationship for phase pressures, leaving only $1$ pressure unknown.
Finally, the Eqn. \eqref{eqn:HenrysLaw} gives a relationship for $\chi_\alpha^{CH_4}$ and Eqn. \eqref{eqn:RaoultsLaw} gives a relationship for $\chi_\alpha^{H_2O}$, thus reducing the unknown mole fractions to $2$. This leaves $7$ primary unknowns which need to be solved for the coupled system which includes
$4$ nonlinear second order PDEs \eqref{eqn:massBalance_CH4_H2O}, \eqref{eqn:massBalance_salt}, and \eqref{eqn:energyBalance}, $1$ nonlinear nonhomogeneous first order ODE \eqref{eqn:massBalance_hydrate}, and $2$ inequality constraints \eqref{eqn:complementaryConditions}.

\paragraph{}
We choose the following set of primary variables,
\begin{align}
 \label{eqn:primaryVariables}
 \mathcal{P} \defas \left( P_w , S_g , S_h , \chi_w^c , X_w^{CH_4} , X_g^{H_2O} , T \right) \ .
\end{align}
This choice of primary variables is not unique, and depends on the actual application.
In our case, the applications of interest arise from marine geological settings where gas phase may or may not exist, and along with the hydrate phase saturations, the gas phase saturation and dissolved methane mole fraction are the most important quantities of interest, and therefore, \eqref{eqn:primaryVariables} is the most suitable choice.

\section{Numerical Solution Strategy}
\label{sec:numerical_solution}
\subsection{Space and time discretization of the conservation laws}
\label{subsec:discretization}

\paragraph{}
The Eqns. \eqref{eqn:massBalance_CH4_H2O}-\eqref{eqn:massBalance_salt}, and \eqref{eqn:energyBalance} are discretized in space using a classical cell-centered finite volumes method defined on orthogonal meshes $T_h$ with $\mathcal{N}$ finite volume cells. 
The fluxes are evaluated using a two-point finite difference approximation of the gradients. 
Convective fluxes are fully upwinded. 
For time discretization, an implicit Euler method is used.
The details of the discretization scheme can be found in \cite{SGuptaThesis2016}.

\paragraph{}
The discretized model can be represented as a system of nonlinear algebraic equations as,
\begin{align}
 \label{eqn:discretized_model}
 \mathbf{F} \defas \mathbf{A}\left( \mathbf{X}^{n+1},\mathbf{X}^{n} \right) \mathbf{X}^{n+1} + \mathbf{B}\left( \mathbf{X}^{n+1},\mathbf{X}^{n} \right) = \mathbf{0} \ ,
\end{align}
where, $\mathbf{X}$ denotes the solution vector which contains the discrete finite volume approximations of the unknowns $\mathcal{P}$ at each cell center.
The indices $n+1$ and $n$ denote the solution at times $t^{n+1}$ and $t^n$. 

\subsection{Nonlinear complementary constraints}
\label{subsec:NCC}

\paragraph{}
The complementarity constraints \eqref{eqn:complementaryConditions} can be rewritten as equivalent non-differentiable but semi-smooth functions as proposed in \cite{Lauser2011},
\begin{align}
 \label{eqn:complementaryConditionsReformulated}
 \qquad S_\alpha - \max\left\{ 0, S_\alpha - \left( 1 - \sum\limits_{\kappa} \chi_\alpha^\kappa \right) \right\} = 0 \ , \quad \forall \alpha \ .
\end{align}
which are piecewise linear with respect to the variables $S_\alpha$, $\chi_\alpha^{\kappa}$.
Such functions are commomnly referred as complementary functions or $NCP-$functions in literature.
Some examples of other forms of such functions include the minimum function and Fischer-Burmeister function (see \cite{Facchinei2013,ChenChenKanzow2000,Fischer1995a,Fischer1995b,Fischer1992}).

\paragraph{}
The complementarity constraints \eqref{eqn:complementaryConditions} and their equivalent form \eqref{eqn:complementaryConditionsReformulated} are local in nature, and must hold cell-wise as,
\begin{align}
 \label{eqn:NCCCellWise}
 \forall j \in \mathcal{N}: \qquad \left(S_\alpha\right)_j - \max\left\{ 0, \left(S_\alpha\right)_j - \left( 1 - \sum\limits_{\kappa} \left( \chi_\alpha^\kappa \right)_j \right) \right\} = 0 \ , \quad \forall \alpha \ .
\end{align}
Note, the degrees of freedom of \eqref{eqn:NCCCellWise} can be partitioned into the following active-inactive sets:
\begin{align}
 \label{eqn:active_inactive_sets}
 \mathcal{A}_\alpha \defas \left\{ j \in \mathcal{N} : \left(S_\alpha\right)_j - \left( 1 - \sum\limits_{\kappa} \left( \chi_\alpha^\kappa \right)_j \right) > 0 \right\} \ , 
 \qquad \mathcal{I}_\alpha \defas \mathcal{N} \backslash \mathcal{A}_\alpha \ .
\end{align}
The active sets $\mathcal{A}_\alpha$ corresponds to the cells where phase $\alpha$ is present, while the inactive sets $\mathcal{I}_\alpha$ correspond to the calls where phase $\alpha$ is absent.

\paragraph{}
Using relationships \eqref{eqn:summationcondition_phases}, \eqref{eqn:HenrysLaw}, and \eqref{eqn:RaoultsLaw} in Eqn. \eqref{eqn:NCCCellWise}, we get the following system of non-differentiable equations,
\begin{align}
 \label{eqn:NCC-1}
 &
 \mathcal{C}_g \left(\mathcal{P}_j^{n+1}\right) \defas \left(S_g\right)_j^{n+1} - \max\left\{ 0, \left(S_g\right)_j^{n+1} - \left( 1 - {\Pi_g}_j^{n+1} \left( \chi_w^{CH_4} \right)_j^{n+1} - \left( \chi_g^{H_2O} \right)_j^{n+1}  \right) \right\} = 0 \\
 \label{eqn:NCC-2}
 & 
 \mathcal{C}_w \left(\mathcal{P}_j^{n+1}\right)  \defas 1 - \left(S_g\right)_j^{n+1} - \left(S_h\right)_j^{n+1} \notag \\  
 &
    - \max\left\{ 0, 1 - \left(S_g\right)_j^{n+1} - \left(S_h\right)_j^{n+1} -         \left( 1 - \left( \chi_w^{CH_4} \right)_j^{n+1} - {\Pi_w}_j^{n+1}\left( \chi_g^{H_2O} \right)_j^{n+1} - \left( \chi_w^c \right)_j^{n+1} \right) \right\} = 0 
\end{align}
where, ${\Pi_g} := \dfrac{H_w^{CH_4}}{z^{CH_4}P_g}$ and ${\Pi_w} := \dfrac{P_g}{P_{sat}^{H_2O}}$.

\subsection{Semismooth Newton scheme}
\label{subsec:semismoothNewton}

\paragraph{}
The system of equations \eqref{eqn:NCC-1}-\eqref{eqn:NCC-2} is semi-smooth and piecewise differentiable.
We solve these equations together with the system \eqref{eqn:discretized_model} within the same iterative loop using a generalized variant of the Newton scheme for semi-smooth problems \cite{HagerWohlmuth2010}.
The classical Newton method is valid in all regions where the Eqns. \eqref{eqn:NCC-1}-\eqref{eqn:NCC-2} are differentiable, while in other regions where the Eqns. \eqref{eqn:NCC-1}-\eqref{eqn:NCC-2} are non-differentiable, the Jacobian can be evaluated by extending the value of the derivatives from the neighbourhood of the non-differentiable regions.

\paragraph{}
We approximate the Jacobian for our Newton scheme using a central difference method. 
To approximate the Jacobian for the Eqns. \eqref{eqn:NCC-1}-\eqref{eqn:NCC-2}, due to their piecewise smooth nature, we use the approximate active/inactive sets $\mathcal{A}_\alpha^{(l)}$ and $\mathcal{I}_\alpha^{(l)}$ at the $l$-th Newton step to determine the phase wise NCP equations in each cell,
\begin{align}
  \label{eqn:NCC_l_1}
    \mathcal{C}_g^{(l)} &=
    \left\{\begin{array}{lr}
    1 - {\Pi_g}_j^{n+1} \left( \chi_w^{CH_4} \right)_j^{n+1} - \left( \chi_g^{H_2O} \right)_j^{n+1}, & \text{for } j \in \mathcal{I}_g^{(l)}\\
    \left(S_g\right)_j^{n+1}, & \text{for } j \in \mathcal{A}_g^{(l)}
    \end{array}\right.
  \\
  \label{eqn:NCC_l_2}
    \mathcal{C}_w^{(l)} &=
    \left\{\begin{array}{lr}
    1 -\left( \chi_w^{CH_4} \right)_j^{n+1} - {\Pi_w}_j^{n+1}\left( \chi_g^{H_2O} \right)_j^{n+1} - \left( \chi_w^c \right)_j^{n+1}, & \text{for } j \in \mathcal{I}_w^{(l)}\\
    1 - \left(S_g\right)_j^{n+1} - \left(S_h\right)_j^{n+1}, & \text{for } j \in \mathcal{A}_w^{(l)}
    \end{array}\right.  
\end{align}
The approximate active/inactive sets $\mathcal{A}_\alpha^{(l)}$ and $\mathcal{I}_\alpha^{(l)}$ may change several times during the Newton loop, but if the Newton method converges, the final active sets will correspond to the physically correct phase state of the system.
The advantage of our semi-smooth Newton scheme is that the treatment of the phase transitions is consistent within the Newton loop, which makes it easier to determine the physically correct phase state even for strongly coupled phase transitions.
The Newton iteration is rather robust with respect to the initialization of the active/inactive sets, and therefore, larger time step sizes can be used.

\subsection{Numerical implementation}
\label{subsec:numerical_implementation}

\paragraph{}
We implemented the semi-smooth Newton scheme described in Sec.\ref{subsec:semismoothNewton} for solving the nonlinear system \eqref{eqn:discretized_model},\eqref{eqn:NCC-1}-\eqref{eqn:NCC-2} within the DUNE-PDElab framework \cite{Bastian2010} which is based on C++. 
For solution of the linear system arising from the Newton-linearization, we used a SUPERLU linear solver \cite{superLU99_SEQ}.
For parallel computations, we use a parallel algebraic multigrid (AMG) solver which uses a stabilized bi-conjugate gradient method as a preconditioner and a symmetric successive over-relaxation smoothening algorithm. The AMG solver is built-in the dune-istl library (https://www.dune-project.org/modules/dune-istl/).
For making the numerical computations, we used the NEC HPC-Linux-Cluster which is part of a hybrid NEC high performance system at the University Computing Centre of the Christian Albrecht Universit\"at, Kiel, Germany.

\section{Numerical Examples}
\label{sec:numerical_examples}
\paragraph{}
Here, we present three numerical examples.
In Example 1, we validate our numerical model by considering a series of phase transitions involving appearance and disappearance of the gas phase, and comparing the solution with that of a PVS scheme. 
In Example 2, we simulate the sedimentation driven gas migration through the GHSZ in the highly dynamic geological setting of the Black Sea, and compare the performance of our semi-smooth Newton scheme against a PVS scheme to show the robustness of our numerical scheme.
Finally, in Example 3, we simulate a gas production scenario, also based on the geological setting of the Black Sea, where we consider that the hydrate phase is randomly distributed within the hydrate layer with saturations ranging between $0-0.6$, and show that our numerical scheme can robustly handle phase transitions even when the phase distributions and the permeability and porosity profiles are highly complex with large variations.

\subsection{Example 1: Model validation}
\label{subsec:example1}
\paragraph{}
In this example, we verify our numerical scheme by simulating a series of phase transitions over time and comparing the solution with that of a PVS scheme.
% In this example, we verify our numerical scheme against a PVS scheme by simulating a series of phase transitions over time and comparing the time at which the gas phase appears and then disappears for the schemes.
We start with zero free gas in the domain, and simulate, by manipulating the system pressure, the appearance of the gas phase as a result of hydrate dissociation, followed by the disappearance of the gas phase due to a combination of hydrate reformation and methane dissolution. 
% If both the schemes can predict the same time at which the gas phase appears and then disappears, then this implies that 

\subsubsection{Problem setting}
\paragraph{}
We consider a $1m \times 1m$ domain $\Omega$ discretized into $50 \times 50$ finite volume cells, and denote the boundary of this domain by $\partial \Omega$. 
At $t=0$, only the hydrate and the aqueous phases are present in the domain. The gas phase is not present (i.e., $\left.S_g\right|_{t=0}=0$), and the aqueous phase contains no dissolved methane (i.e., $\left.\chi_w^{CH_4}\right|_{t=0}=0$). 
The hydrate phase is uniformly distributed throughout the domain and has an initial saturation of $\left.S_h\right|_{t=0}=0.3$.
The initial concentration of the dissolved salt is $\left.\chi_w^{c}\right|_{t=0}=5.5$ mmol/mol of water, and the initial temperature in the domain is $\left.T\right|_{t=0}=4^oC$.
At the initial temperature, pressure, and salinity conditions, the hydrate equilibrium pressure is $\left.P_{eq}\right|_{t=0} = 3.4 \text{ MPa}$, which is higher than the initial gas pressure $\left.P_{g}\right|_{t=0} = 2.0848 \text{ MPa}$. So, the hydrate is in an unstable state.

\paragraph{}
For all $t>0$, the temperature at the boundary is maintained at the initial value, i.e., $\left.T\right|_{\partial\Omega}=4^oC$,
and a zero-gas-flux condition is prescribed at the boundary, i.e., $\left.\mathbf{v}_g\right|_{\partial\Omega}=0$.
The phase transitions are triggered by manipulating the boundary conditions for $P_w$.

\paragraph{} 
The initial and the boundary conditions are summarized in Table \ref{table:p1_ICBCs}.
The relevant material properties are listed in Table \ref{table:material_properties_and_model_parameters}.
The problem setting is chosen such, that the spatial variations across the domain are negligible, and the focus of the problem remains on the phase transitions.

\subsubsection{Phase transitions}
\paragraph{}
The manipulation of the boundary conditions for $P_w$ corresponds to the following four stages (refer Fig. \ref{fig:p1_Pg_over_t}):
\begin{enumerate}
 \item For the period $0 < t < 200 \text{ hrs}$, $\left.P_w\right|_{\partial\Omega}$ is held constant at $2$ MPa. 
 Since hydrate is unstable at this pressure, it will dissociate to produce $CH_4$, which will dissolve into the porewater as long as $\chi_w^{CH_4} < \chi_{w,sat}^{CH_4}\left( P_w, T, X_c\right) $, where, $\chi_{w,sat}^{CH_4}$ refers to the solubility of methane in the aqueous phase. 
 If the solubility is reached, i.e., $\chi_w^{CH_4} = \chi_{w,sat}^{CH_4}\left( P_w, T, X_c\right) $, the gas phase will appear.
 \item Next, for the period $200 \text{ hrs} \leq t < 350 \text{ hrs}$, a zero-water-flux condition is prescribed, i.e., $\left.\mathbf{v}_w\right|_{\partial\Omega}=0$, s.t., the domain is now fully closed. 
 Under these conditions, hydrate will continue to dissociate and pore-pressures will rise until a state of equilibrium is reached s.t., $P_g=P_{eq}$.
 \item At $t=350 \text{ hrs}$, $\left.P_w\right|_{\partial\Omega}$ is instantaneously stepped-up to $5$ MPa, and this pressure is maintained for the period $350 \text{ hrs} < t < 450 \text{ hrs}$.
 Due to a step increase in the pressure and following the VLE assumption, the gaseous methane will dissolve instantaneously into the porewater, and the gas phase may or may not disappear, depending on how high the new solubility is.
 If the gas phase does not fully disappear at $t=350 \text{ hrs}$, the remaining methane in the gas phase will react with the porewater to form hydrate until the gas phase vanishes.
 \item Finally, for $t\geq 450 \text{ hrs} $, $\left.P_w\right|_{\partial\Omega}$ is linearly ramped up at a rate of $10$ Pa/s.
 If the gas phase is still present at $t=450 \text{ hrs}$, methane dissolution as well as hydrate formation will continue until the gas phase vanishes and only the aqueous and the hydrate phases remain.
\end{enumerate}

\subsubsection{Numerical simulation and results}
\label{subsubsec:p1_numerical_simulation_and_results}

\paragraph{}
The numerical simulation was run until $t_{end}=600 \text{ hrs}$.
The maximum time step size was chosen as $dt_{max}=3600$ s.
An adaptive time-stepping strategy was used where the time step size is controlled heuristically based on the number of Newton iterations per time integration step. 
If the number of Newton iterations is more than $\ell_{h}$, the time step size for the next time integration step is decreased by $25\%$, whereas, if the number of Newton iterations is less than $\ell_{l}$, the time step size is increased by $10\%$.
Between $\ell_{l}$ and $\ell_{h}$ Newton iterations, $dt$ is not changed.
The choice of $\ell_{l}$ and $\ell_{h}$ depends on the problem setting and, as a rule of thumb, in our numerical scheme we consider $\ell_l \geq 4$ and $\ell_h = \ell_l+4$.
In this example, we chose $\ell_{l}=4$ and $\ell_{h}=8$.

\begin{figure}[tbh]
 \centering
    \subfigure[Gas phase pressure over time.]{
        \includegraphics[width=0.475\textwidth]{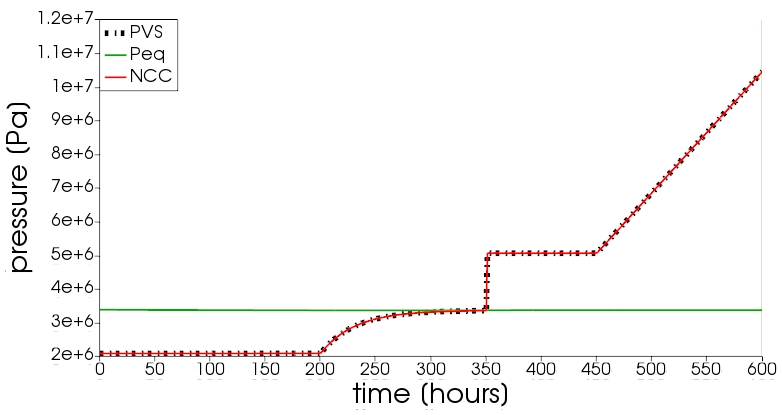}
        \label{fig:p1_Pg_over_t}
    }
    \hfill
    \subfigure[Gas phase saturation over time.]{
        \includegraphics[width=0.475\textwidth]{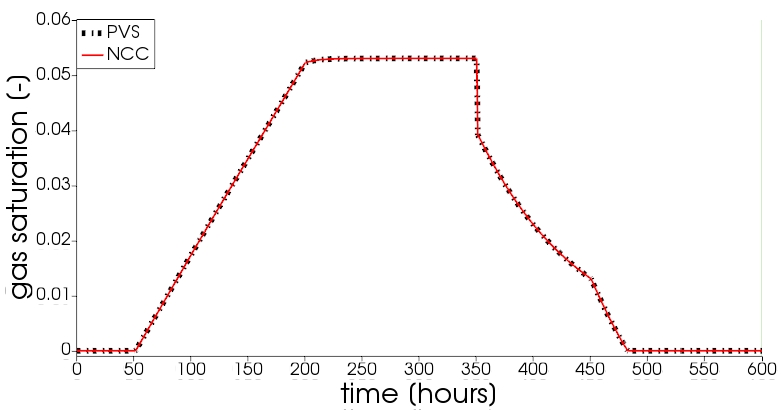}
        \label{fig:p1_Sg_over_t}
    }
    \vfill
    \subfigure[Dissolved $CH_4$ mole-fraction over time.]{
        \includegraphics[width=0.475\textwidth]{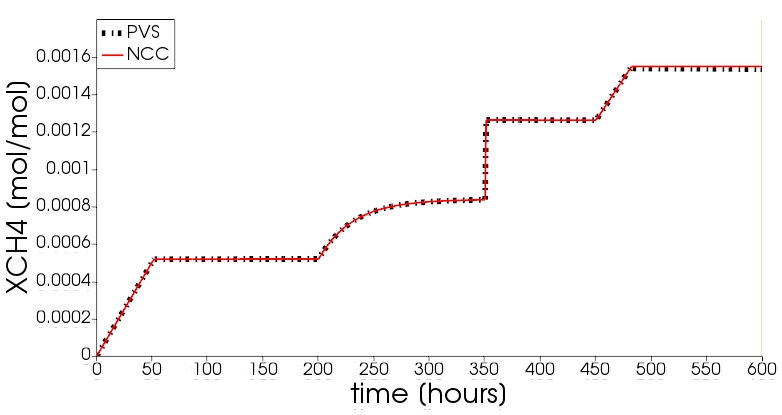}
        \label{fig:p1_XCH4_over_t}
    }
    \hfill
    \subfigure[Phase state over time (PVS).]{
        \includegraphics[width=0.475\textwidth]{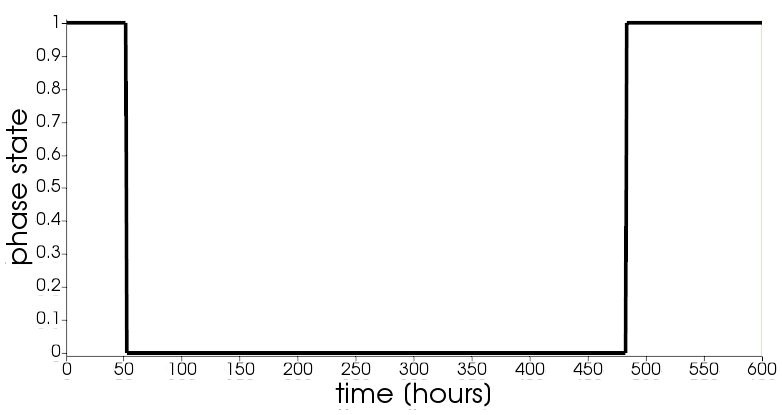}
        \label{fig:p1_phase_state_over_t}
    }
 \caption{Numerical results for the Example 1 (Sec. \ref{subsec:example1}).}
 \label{fig:p1_results}
\end{figure}

\paragraph{}
The numerical results are shown in Fig. \ref{fig:p1_results}, where the $P_w$, $S_g$, $\chi_w^{CH_4}$ profiles and the phase state at the point $\left(0.5m,0.5m\right)$ are plotted over time. 
For the phase state, a value of $0$ indicates that the gas phase is present, while a value of $1$ indicates that the gas phase is absent.
The results show that in the first stage, the gas phase appears at $t=52 \text{ hrs}$.
Between $52 \text{ hrs} < t < 200 \text{ hrs}$, $S_g$ increases as the hydrate continues to dissociate (Fig. \ref{fig:p1_Sg_over_t}).
In the second stage, an equilibrium state is achieved at $t\approx 330 \text{ hrs}$, s.t., between $330 \text{ hrs} < t < 350 \text{ hrs} $ no further gas dissolution and hydrate phase changes occur (Figs. \ref{fig:p1_Sg_over_t},\ref{fig:p1_XCH4_over_t}).
In the third stage, at $t=350 \text{ hrs}$, solubility of methane is too small for the gas phase to vanish. 
Between $350 \text{ hrs} < t < 450 \text{ hrs}$, the pressure is constant at $5$ MPa, and a steady state is reached for the dissolved methane, i.e. the rate of gas dissolution equals the rate of hydrate formation, s.t., $S_g$ decreases as hydrate formation continues (Fig. \ref{fig:p1_Sg_over_t}), while $\chi_w^{CH_4}$ remains constant (Fig. \ref{fig:p1_XCH4_over_t}).
At $t=450 \text{ hrs}$, gaseous methane is still present. As the pressure is ramped up in the fourth stage, both hydrate formation as well as gas dissolution continue until the gas phase finally disappears at $t=482 \text{ hrs}$. 

\paragraph{}
In order to ensure that our implementation of the NCP approach for the phase transitions is correct, we compare our results with the more common primary variable switching (PVS) approach of \cite{Class2002}. 
We implemented this PVS scheme within the same software framework as our NCP approach, i.e., DUNE-PDElab, version 2.6.0 (https://www.dune-project.org/modules/dune-pdelab/).
For both the schemes, we used the same discretization scheme (Sec. \ref{subsec:discretization}) and the same linear solver (SuperLU).
For the Newton solver, we used the same convergence criteria, and for the adaptive time stepping strategy, we used identical control parameters.
Note that, we implemented only a sequential version of the PVS scheme. Therefore, for those examples where we compare the solution of our NCP scheme with the PVS scheme, we performed all numerical simulations only in a seuential mode. 

\paragraph{}
We can see in Fig. \ref{fig:p1_results} that both the NCP and the PVS schemes are in agreement over the predicted sequence of the $hydrate \leftrightharpoons gaseous\ methane \leftrightharpoons dissolved\ methane$ phase transitions.

\begin{table}[tbh]
 \caption{Initial and bounday conditions for the Example 1 (Sec. \ref{subsec:example1}).}
 \label{table:p1_ICBCs}
 \centering
    \begin{tabular}{| r !{\color{lightgray}\vrule} r c l |}
    \hline
    \multicolumn{4}{|c|}{Initial conditions} \\
    \hline
    \multirow{6}{*}{at $t=0$ , and $\mathbf{x} \in \Omega$} 
    & $P_w$             & $ = $ & $ 2 \text{ MPa}$           \\
    & $S_g $            & $ = $ & $ 0 $                     \\ 
    & $S_h $            & $ = $ & $ 0.3 $                   \\ 
    & $\chi_w^{CH_4} $  & $ = $ & $ 0 $                     \\ 
    & $\chi_g^{H_2O} $  & $ = $ & $ \chi_{g,sat}^{H_2O}\left( \left.P_g\right|_{t=0}, \left.T\right|_{t=0} \right) $   \\ 
    & $\chi_w^c $       & $ = $ & $ 5.5 \text{ mmol/mol} $   \\
    & $T $              & $ = $ & $ 4^oC $                  \\
    \hline
    \multicolumn{4}{|c|}{Boundary conditions} \\ 
    \hline
    \multirow{4}{*}{for $\mathbf{x} \in \partial\Omega$, \qquad \quad }
    $0 < t < 200 \text{ hrs}$                   & $P_w$           & $=$ & $2\text{ MPa}$   \\
    $200 \text{hrs} \leq t < 350 \text{ hrs}$   & $\mathbf{v}_w$  & $=$ & $0$             \\
    $350 \text{hrs} \leq t < 450 \text{ hrs}$   & $P_w$           & $=$ & $5\text{ MPa}$   \\
    $t \geq 450 \text{ hrs}$                    & $P_w$           & $=$ & $5\text{ MPa} + 10.\times\left( t - 450\times 3600\right)$   \\ 
    \arrayrulecolor{lightgray}\hline\arrayrulecolor{black}
    \multirow{3}{*}{for $\mathbf{x} \in \partial\Omega$, and $t>0$ }
    & $\mathbf{v}_g$    & $=$ & $0$     \\
    & $\nabla \chi_w^c$ & $=$ & $0$     \\
    & $T$               & $=$ & $4^oC$  \\
    \hline
    \end{tabular}
\end{table}

\subsection{Example 2: Gas migration through gas-hydrate stability zone (GHSZ)}
\label{subsec:example2}
\paragraph{}
In this example, we simulate the gas hydrate dynamics driven by the changes in temperature, pressure, and salinity conditions as a result of the sediment deposition in the highly dynamic geological setting of the Black Sea.
Through this field-scale environmental application, we aim to demonstrate the complexities and challenges associated with the highly coupled phase transitions in natural gas hydrate systems, and show the robustness of our semi-smooth Newton scheme in realistic settings.
We also compare the performance of our semi-smooth Newton scheme with that of a primary variable switching scheme.

\subsubsection{Problem setting}
\label{subsubsec:example2_problem_setting}

\begin{figure}
 \includegraphics[width=\textwidth]{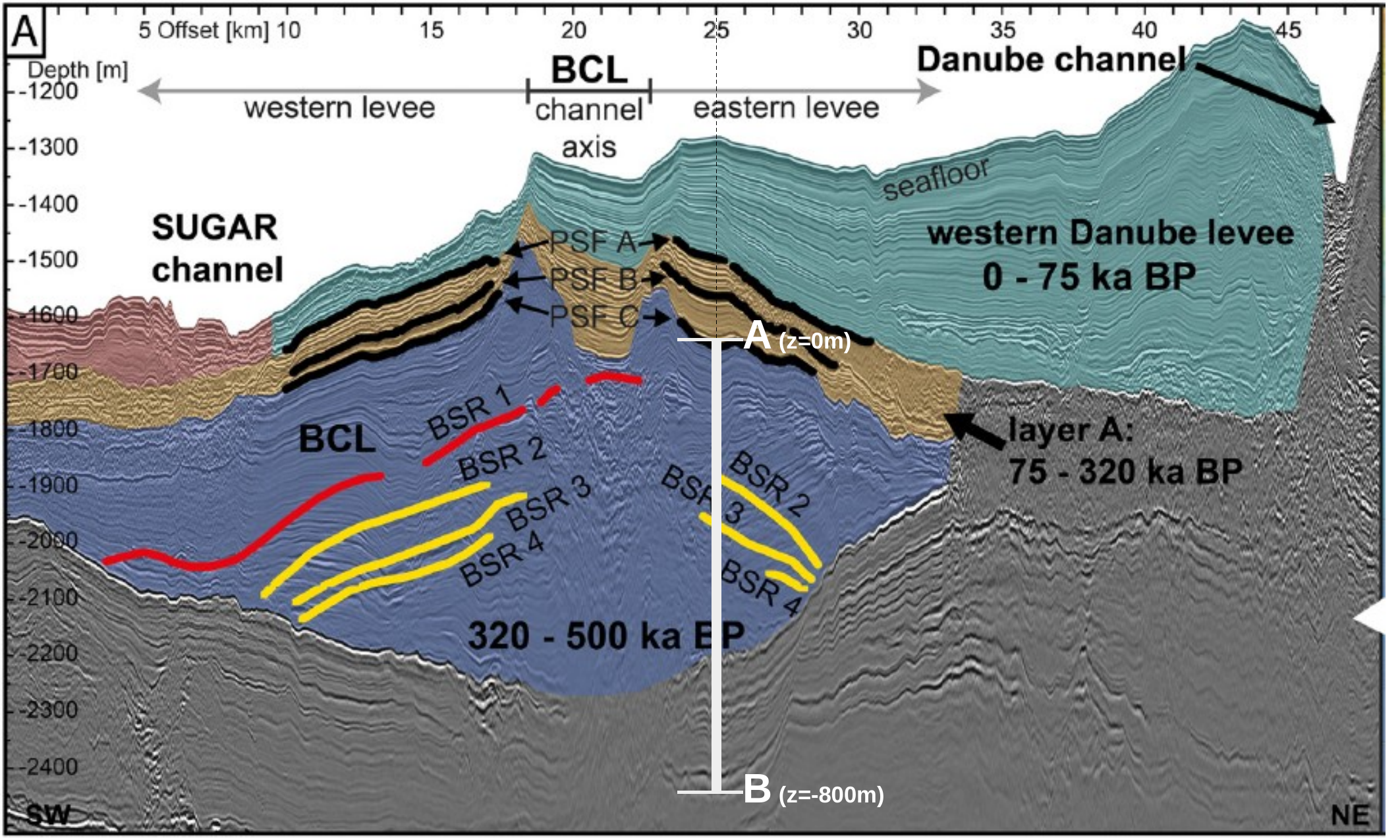}
 \caption{  Regional seismic profile across the western part of the Danube paleo delta in SW to NE direction, depicting the geological setting for Example 2 (Sec.\ref{subsec:example2}).
            2D RMCS line 09.
            Interpretation of the seismic data according to \cite{Zander2017}.}
 \label{fig:p2_geological_setting}
\end{figure}

\paragraph{}
The geological setting for this problem is based on the Danube paleo delta which consists of stacked channel-levee systems that were active during glacial times when the water level was approximately $100-150$ m lower than today \cite{Winguth2000}. 
For our problem, we have chosen a buried channel-levee (BCL) complex (blue color, Fig. \ref{fig:p2_geological_setting}) west of the Viteaz canyon, the main Danube paleo channel, which has buried the BCL over the past $75$ ka (green color, Fig. \ref{fig:p2_geological_setting}). 
The BCL is believed to have deposited its levees essentially in two main events correlating to oxygen isotope stages $8$ and $6$ \cite{Winguth2000}, i.e. between $320$ ka and $75$ ka BP (brown color, Fig. \ref{fig:p2_geological_setting}; \cite{Zander2017}). 
These two active phases of the BCL correspond to limnic stages of the Black Sea that have been documented, for example by low sulfur contents, in the sedimentary record of DSDP drill Site 379 in the eastern basin of the Black Sea \cite{DegensRoss1974}. 
For the past $500$ ka, this drill core identifies five marine stages that are interrupted by four limnic stages, i.e. intervals when the sea level dropped below the depth of Bosphorus sill (today at $40$ m water depth), thereby, separating the Black Sea from saltwater inflow from the Marmara and Mediterranean Sea.

\paragraph{}
In our problem, we are interested in simulating how the deposition of the brown and green sediment layers affects the gas hydrate stability zone (GHSZ) that was established $300$ ka BP, i.e. in the blue sediments (Fig. \ref{fig:p2_geological_setting}).

\paragraph{}
Hence, we base the initial setting $\left(\text{at time } t=t_0=0\right)$ on the paleo conditions existing at $300$ ka BP.
We choose an arbitrary $1D$ segment $A-B$ located in the eastern levee as our computational domain (See Fig.\ref{fig:p2_geological_setting}).
Point $A$ of the computational domain corresponds to the paleo seafloor at $300$ ka BP (PSF-C), i.e., $z=z_A=0$m, where $z$ denotes the depth below the sea floor.
Point $B$ is located at $z=z_B=-800$m.

\begin{figure}
 \includegraphics[width=\textwidth]{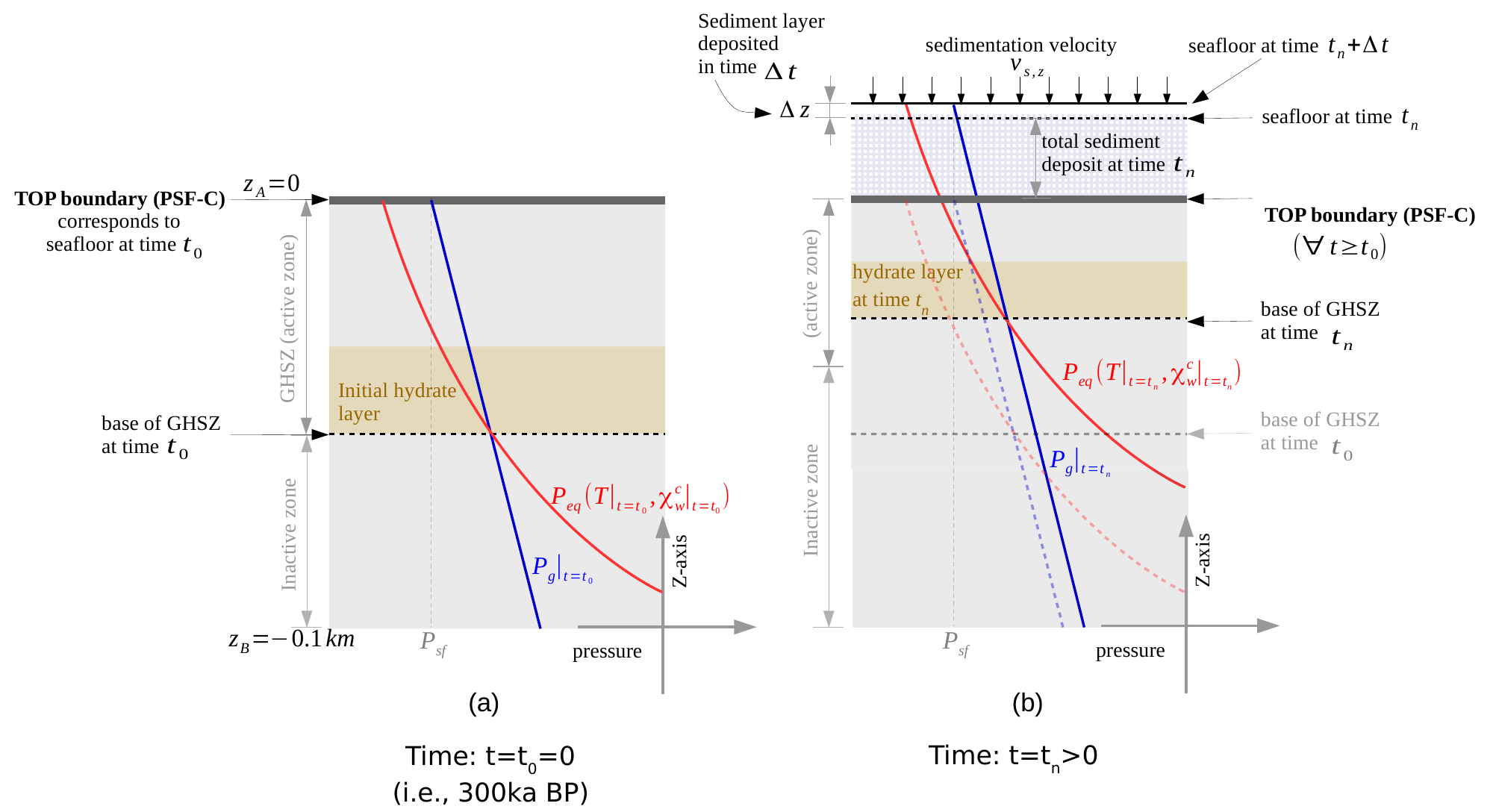}
 \caption{Problem setting for Example 2 (Sec.\ref{subsec:example2}).
          Figure (a) shows the initial state of the system and identifies the corresponding GHSZ. 
          Figure (b) shows the state of the system at $t=t_n>0$ and illustrates how the GHSZ shifts as a result of sedimentation over time.}
 \label{fig:p2_problem_setting}
\end{figure}

\paragraph{}
At $t=t_0$, we assume a hydrostatic pressure at point A of $\left.P_w\right|_{z=0,t=0} = P_{sf} = 15$ MPa, corresponding to a water depth of roughly $1500$m, and a bottom water temperature of $\left.T\right|_{z=0,t=0} = T_{sf} = 4^0$C, corresponding to glacial conditions in the Black Sea \cite{Zander2017}. 
We assume that the initial pressure distribution within the computational domain follows a hydrostatic gradient, and the initial temperature distribution follows a steady state geothermal gradient of $35^0C/km$.
The initial conditions for all the primary variables are listed in Table \ref{table:p2_ICBCs}.
Based on the initial pressure, temperature, and salinity conditions, we can estimate the location of the base of the GHSZ (bGHSZ), i.e., the point of intersection of the gas phase pressure and the gas hydrate equilibrium pressure curves plotted along the sediment depth. 
The gas hydrates are stable above bGHSZ where $P_g \geq P_{eq}$, and unstable below. 
(See Fig.\ref{fig:p2_problem_setting}-a.)
For this setting, the initial bGHSZ is located $400$m below point A,
and we assume that a hydrate layer of $80$m thickess and $30\%$ peak saturation is located just above this initial bGHSZ. 

\paragraph{}
For the sake of simplicity, we assume that the deposition of the brown, and green sediment layers occurs continuously over $300,000$ years at a constant sedimentation rate $v_{s,z}=0.1$ cm/year.
This does not reflect the true depositional history, but rather, simulates a scenario of a low average sedimentation rate.
Fig.\ref{fig:p2_problem_setting}-b shows schematically how the sedimentation shifts the GHSZ.
Basically, due to the sedimentation process, the sea floor rises over time.
At any time $t=t_n>t_0$, the corresponding sea floor PSF-n is located at $z=z_{\text{PSF-n}}= z_A + v_{s,z} t_n$, and within a time increment $\Delta t$, a new sediment layer of thickness $\Delta z = v_{s,z} \Delta t $ is deposited on top of PSF-n.
We assume that $\Delta t$ is small enough for temperature and pressure to reach a steady state within the new sediment layer.
The pressure and temperature at any sea floor PSF-n are are fixed at  
$\left.P_w\right|_{z_{\text{PSF-n}},t_n} = P_{sf}$ and $\left.T\right|_{z_{\text{PSF-n}},t_n} = T_{sf}$, respectively. 
Note that here we ignore any changes in sea level and bottom water temperature during the geological past.
Due to the sedimentation, the temperature and pressure at the top boundary of our computational domain, i.e., point A at $z=0$m, increase over time, which in turn shifts the base of the GHSZ upwards.
(Refer to Table \ref{table:p2_ICBCs} for a list of the boundary conditions, and Table \ref{table:material_properties_and_model_parameters} for a list of material properties and parameters.)
 
\paragraph{}
The main challenge in simulating this setting is that, as the hydrate layer gets buried below the GHSZ due to ongoing sedimentation, it starts to dissociate from the bottom, and the gas phase appears in a narrow region bel thow the GHSZ.
The saturation of the free gas phase is very small, typically less than $5\%$.
The gas migrates upwards due to its buoyancy, but since the overlying hydrate layer has a much lower permeability, the free gas tends to pool below the region where the hydrate stauration is the highest, thereby building up the pore pressure.
The local dilution of the pore water salinity due to fresh water release and the local cooling effect due to hydrate dissociation also give strong feedbacks to both, the hydrate equilibrium pressure, as well as the solubility of the gas in the aqueous phase. 
These competing effects often cause the mathematical model to rapidly switch back and forth between single phase model and two-phase model with respect to the $CH_4 - H_2O$ system, especially when the gas phase appears in the domain for the first time.

\subsubsection{Numerical simulation and results}
\label{subsubsec:example1_numerical_simulation_and_results}

\paragraph{}
The computational domain was discretized uniformly into $1600$ finite volumes along the Z-axis.
The maximum time step size was chosen as $dt_{max}=10$ years,
and the time step size $dt$ was controlled adaptively using the heuristic time stepping strategy described in Sec. \ref{subsubsec:p1_numerical_simulation_and_results} with $\ell_{l}=8$ and $\ell_{h}=12$.

\paragraph{}
We performed the numerical simulations with our semi-smooth Newton (NCP) scheme, and for comparison, also with a primary variable switching (PVS) scheme as discussed in Example 1 (Sec. \ref{subsec:example1}). 

\begin{figure}[tbh]
 \centering
    \subfigure[Problem-time over CPU-time.]{
        \centering
        \includegraphics[width=0.475\textwidth]{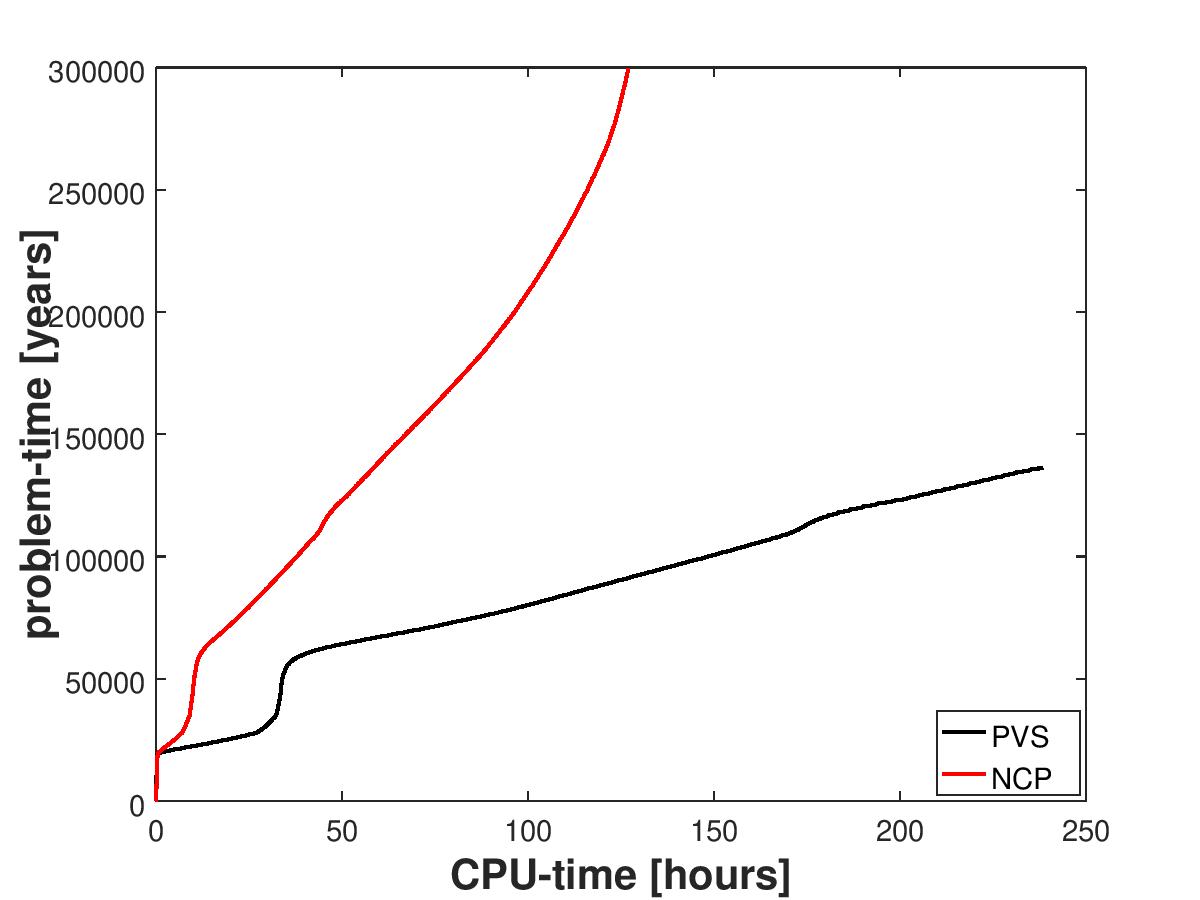}
        \label{fig:p2_cputime_vs_problemtime}
    }
    \hfill
    \subfigure[$dt$ over problem-time.]{
        \centering
        \includegraphics[width=0.475\textwidth]{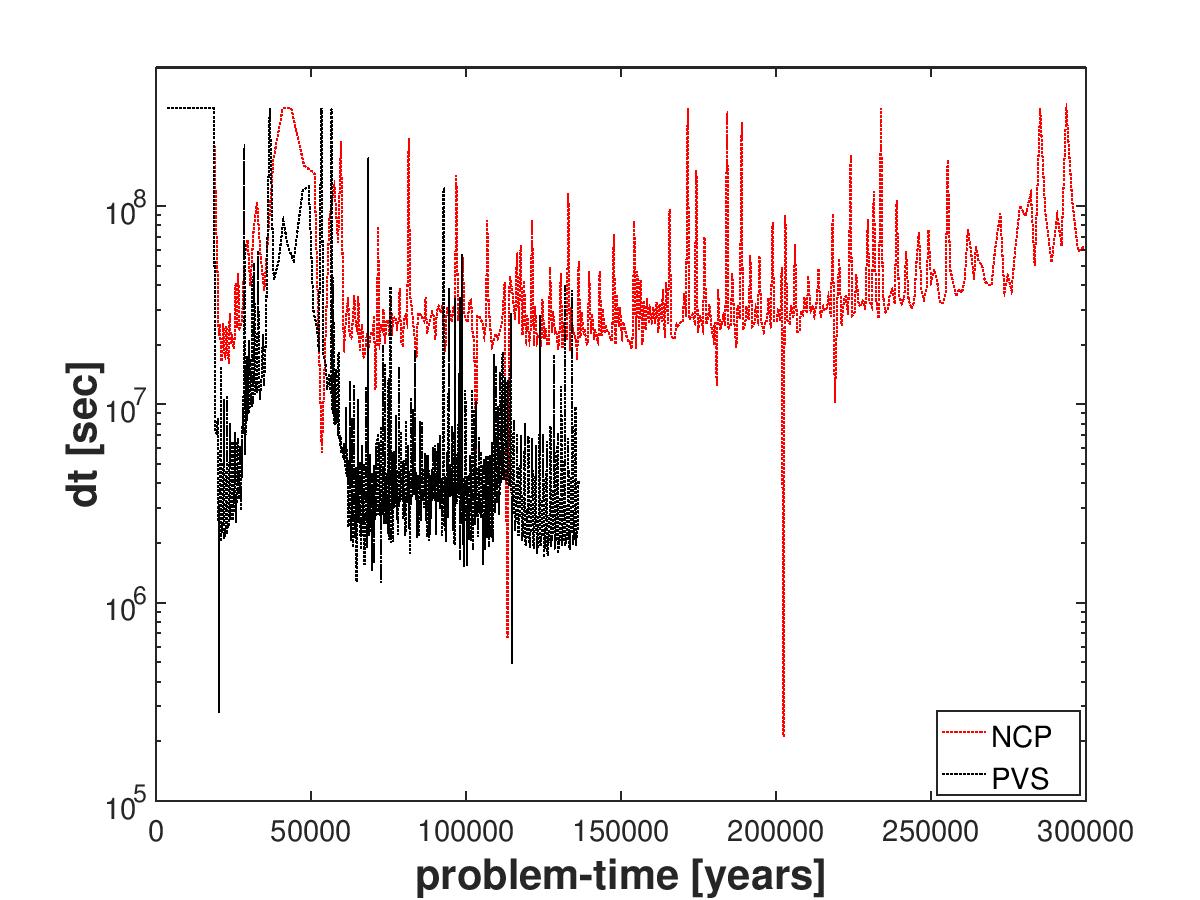}
        \label{fig:p2_problemtime_vs_dt}
    }
    \caption{Numerical results for Example 2 (Sec. \ref{subsec:example2}).
             Figure compares the NCP and the PVS schemes in terms of the cumulative CPU-time required to solve the problem, and the evolution of the time step size during the simulation.}
    \label{fig:p2_results_statistics}
\end{figure}

\paragraph{}
In Fig.\ref{fig:p2_cputime_vs_problemtime}, we can see that the NCP scheme took roughly $120$ CPU-hours to solve the problem upto $t=300,000$ years, whereas, due to drastic reduction in time step size, the PVS scheme could reach only upto $t=135,000$ years in twice as many CPU-hours, which is despite the fact that the PVS scheme needs less time per calculation due to fewer degrees of freedom compared to the NCP scheme. 

\begin{figure}[h]
    \subfigure[$S_h$ and $S_g$ profiles \newline $t=22500$ years]{
        \includegraphics[width=0.3\textwidth]{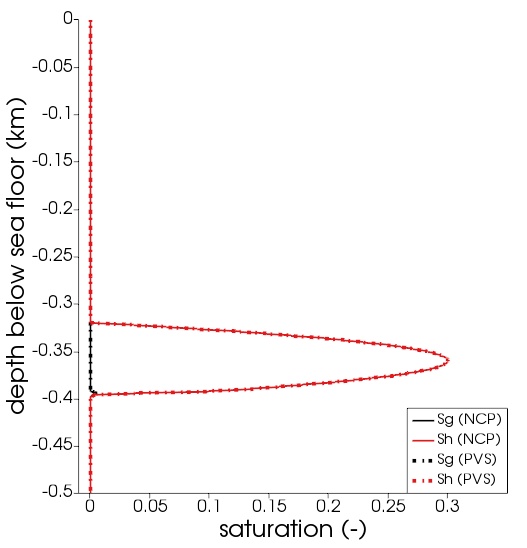}
        \label{fig:p2_Sh_Sg_t=0075}
    }
    \hfill
    \subfigure[$S_h$ and $S_g$ profiles \newline $t=135000$ years]{
        \includegraphics[width=0.3\textwidth]{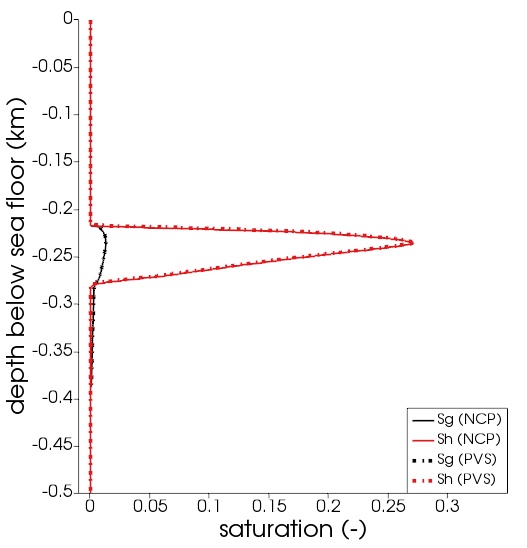}
        \label{fig:p2_Sh_Sg_t=0450}
    }
    \hfill
    \subfigure[$S_h$ and $S_g$ profiles \newline $t=300000$ years]{
        \includegraphics[width=0.3\textwidth]{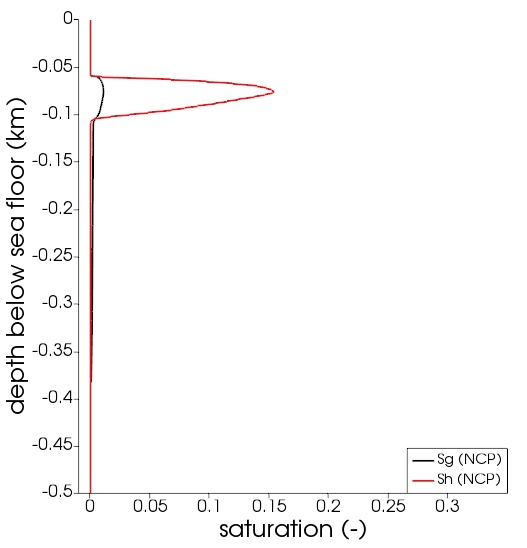}
        \label{fig:p2_Sh_Sg_t=1000}
    }
    \vfill
    \subfigure[$S_g$, $\chi_w^{CH_4}$, and $\chi_w^c$ profiles \newline $t=22500$ years]{
        \includegraphics[width=0.3\textwidth]{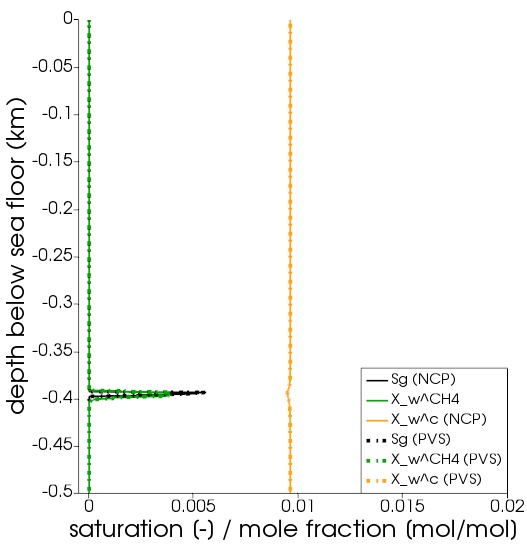}
        \label{fig:p2_Sg_XCH4_Xc_t=0075}
    }
    \hfill
    \subfigure[$S_g$, $\chi_w^{CH_4}$, and $\chi_w^c$ profiles \newline $t=135000$ years]{
        \includegraphics[width=0.3\textwidth]{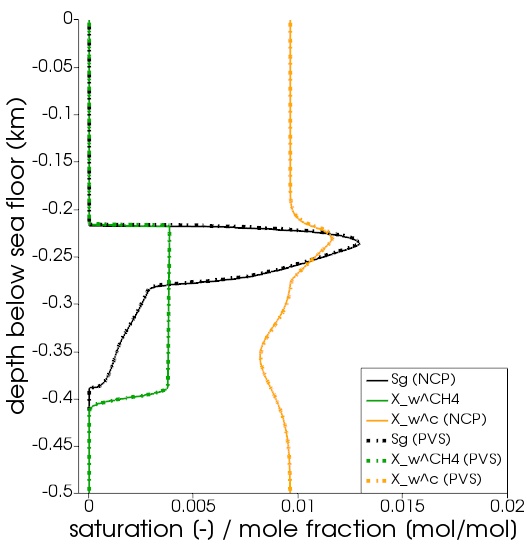}
        \label{fig:p2_Sg_XCH4_Xc_t=0450}
    }
    \hfill
    \subfigure[$S_g$, $\chi_w^{CH_4}$, and $\chi_w^c$ profiles \newline $t=300000$ years]{
        \includegraphics[width=0.3\textwidth]{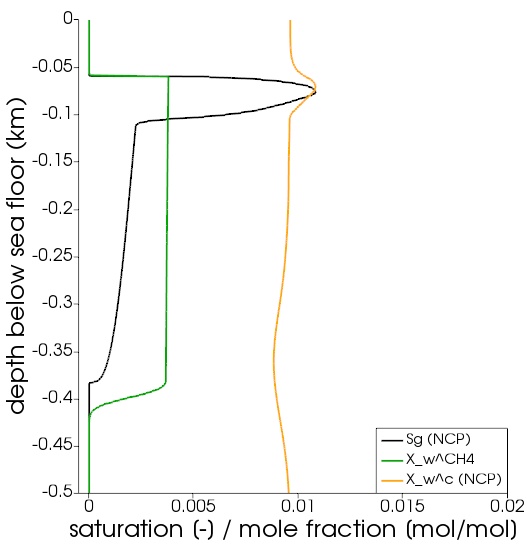}
        \label{fig:p2_Sg_XCH4_Xc_t=1000}
    }
 \caption{Numerical results for Example 2 (Sec. \ref{subsec:example2}).
          Figure shows sapshots of $S_h$, $S_g$, $\chi_w^{CH_4}$, and $\chi_w^c$ at $t=22500$ years, i.e. time when gas phase first appears, $t=135000$ years, i.e. time upto which PVS scheme solved in $240$ CPU-hours, and $t=t_{end}=300000$ years.
          For $t=22500$ years and $t=135000$, the solutions of both NCP and PVS schemes is plotted for comparison.}
 \label{fig:p2_results}
\end{figure}

\begin{figure}[tbh]
    \subfigure[$S_h$, $S_g$ \newline $t=60000$ years]{
        \centering
        \includegraphics[width=0.3\textwidth, angle =270]{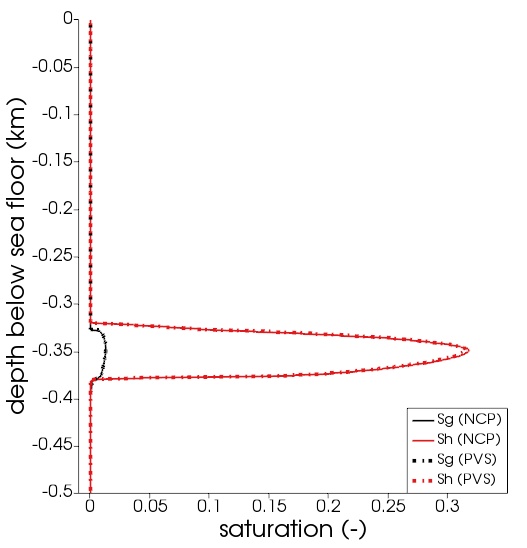}
        \label{fig:hires_p2_Sh_Sg_t=0200}
    }
    \hfill
    \subfigure[$S_g$, $\chi_w^{CH_4}$, $\chi_w^c$ \newline $t=60000$ years]{
        \centering
        \includegraphics[width=0.3\textwidth, angle =270]{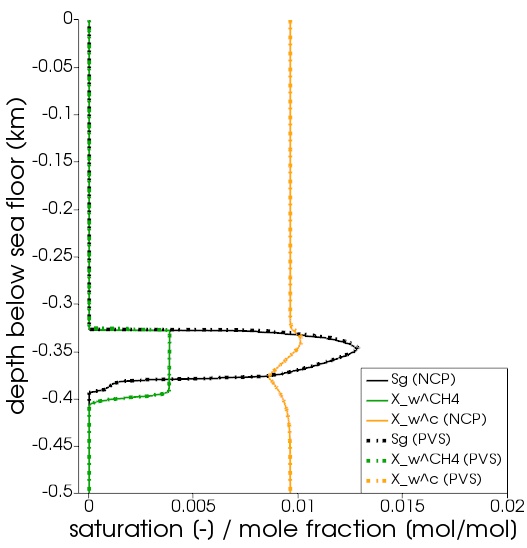}
        \label{fig:hires_p2_Sg_XCH4_Xc_t=0200}
    }
    \vfill
    \subfigure[$S_h$, $S_g$ \newline $t=82500$ years]{
        \centering
        \includegraphics[width=0.3\textwidth, angle =270]{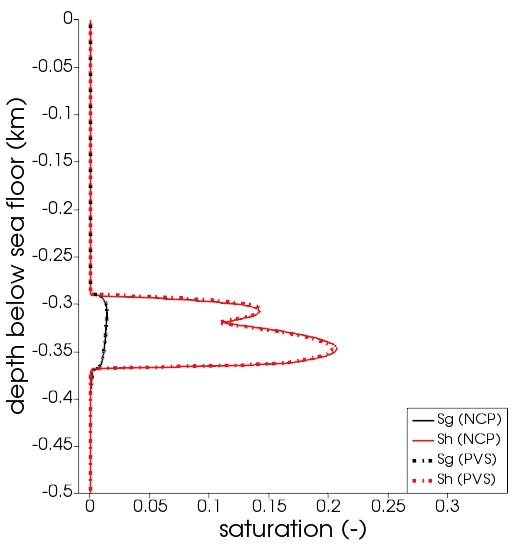}
        \label{fig:hires_p2_Sh_Sg_t=0275}
    }
    \hfill
    \subfigure[$S_g$, $\chi_w^{CH_4}$, $\chi_w^c$ \newline $t=82500$ years]{
        \centering
        \includegraphics[width=0.3\textwidth, angle =270]{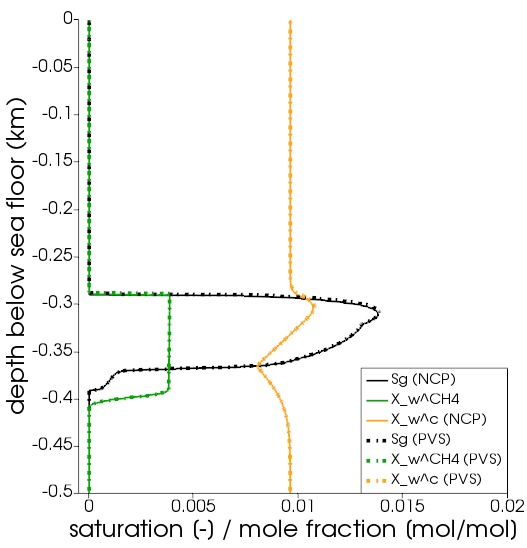}
        \label{fig:hires_p2_Sg_XCH4_Xc_t=0275}
    }
    \vfill
    \subfigure[$S_h$, $S_g$ \newline $t=90000$ years]{
        \centering
        \includegraphics[width=0.3\textwidth, angle =270]{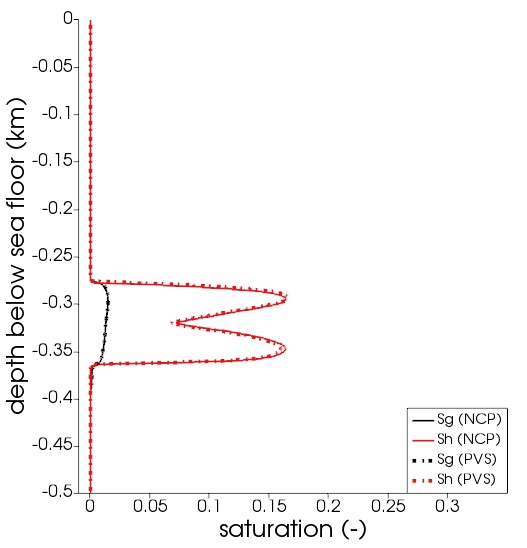}
        \label{fig:hires_p2_Sh_Sg_t=0300}
    }
    \hfill
    \subfigure[$S_g$, $\chi_w^{CH_4}$, $\chi_w^c$ \newline $t=90000$ years]{
        \centering
        \includegraphics[width=0.3\textwidth, angle =270]{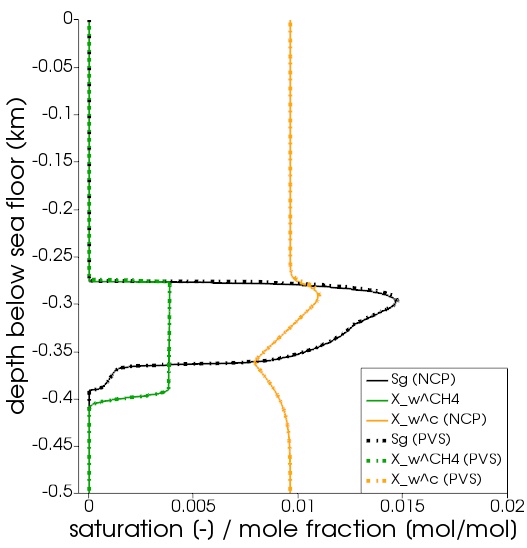}
        \label{fig:hires_p2_Sg_XCH4_Xc_t=0300}
    }
    \vfill
    \subfigure[$S_h$, $S_g$ \newline $t=105000$ years]{
        \centering
        \includegraphics[width=0.3\textwidth, angle =270]{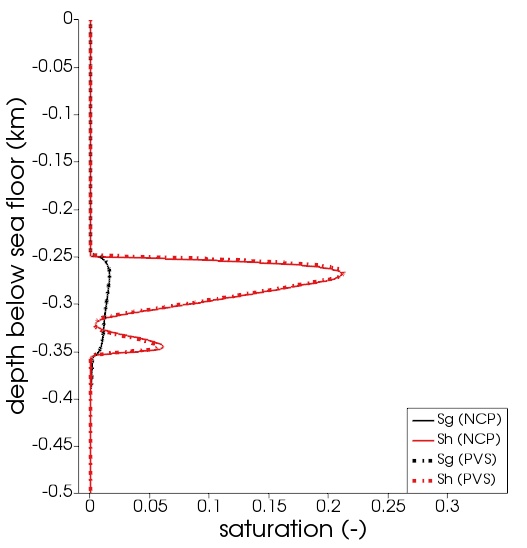}
        \label{fig:hires_p2_Sh_Sg_t=0350}
    }
    \hfill
    \subfigure[$S_g$, $\chi_w^{CH_4}$, $\chi_w^c$ \newline $t=105000$ years]{
        \centering
        \includegraphics[width=0.3\textwidth, angle =270]{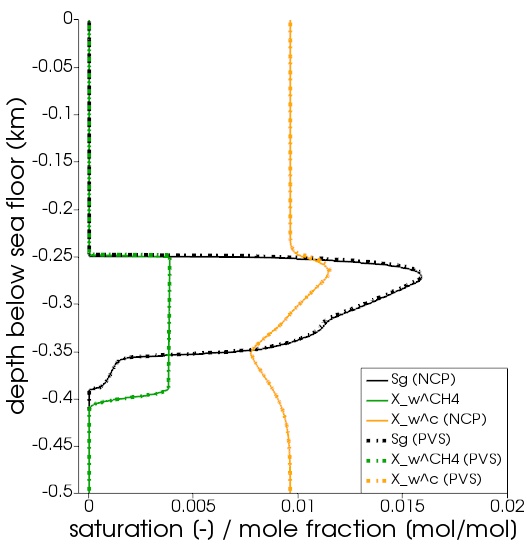}
        \label{fig:hires_p2_Sg_XCH4_Xc_t=0350}
    }
    \vfill
    \subfigure[$S_h$, $S_g$ \newline $t=120000$ years]{
        \centering
        \includegraphics[width=0.3\textwidth, angle =270]{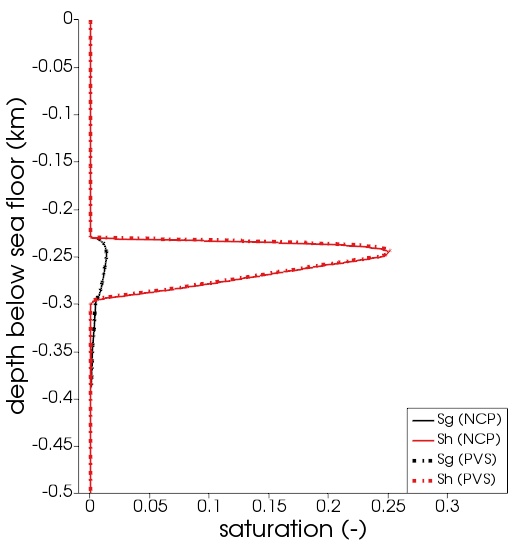}
        \label{fig:hires_p2_Sh_Sg_t=0400}
    }
    \hfill
    \subfigure[$S_g$, $\chi_w^{CH_4}$, $\chi_w^c$ \newline $t=120000$ years]{
        \centering
        \includegraphics[width=0.3\textwidth, angle =270]{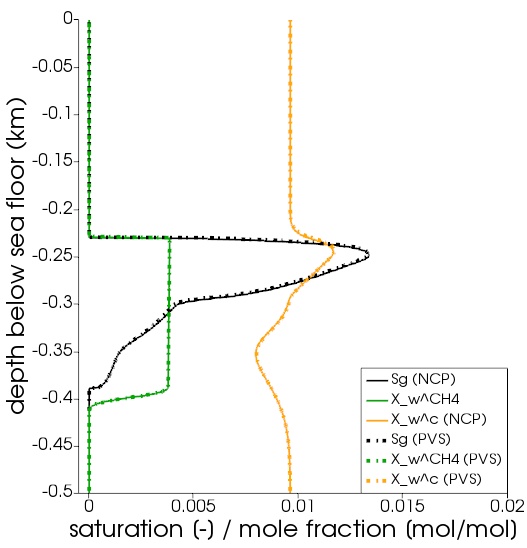}
        \label{fig:hires_p2_Sg_XCH4_Xc_t=0400}
    }
    \caption{Numerical results for Example 2 (Sec. \ref{subsec:example2}).
                Figure shows the process of hydrate dissociation $\rightarrow$ gas migration $\rightarrow$ hydrate reformation as a result of rising GHSZ between $60,000$ years $\leq t \leq$ $120,000$ years. 
                New gas hydrate layer grows using the methane gas supplied by the dissociating gas hydrate layer below.}
    \label{fig:hires_p2_results}
\end{figure}

\paragraph{}
In Fig.\ref{fig:p2_results}, the snapshots of $S_h$, $S_g$, $\chi_w^{CH_4}$, and $\chi_w^c$ are plotted at three times: $t_1=22,500$ years, $t_2=135,000$ years, and $t_3=300,000$ years. 
Time $t_1$ corresponds to the instant when the gas phase first appears in the domain. 
We can see that both the PVS and the NCP schemes are in agreement about where the gas phase appears and in what saturation.
Time $t_2$ corresponds to the time upto which the PVS scheme could solve in $240$ CPU-hours (at which point the simulation was aborted due to large run time).
The results of the PVS and the NCP schemes show a very good match, and serve as a good validation for our numerical implementation.
Time $t_3$ corresponds to the end-time for this problem. Only the solution of the NCP scheme is plotted for this time step.
The base of the GHSZ shifts upwards by $300$m due to sedimentation over a period of $300,000$ years.
The results show that as the GHSZ rises towards the sea floor,
the hydrate layer dissociates, generating methane below the base of the GHSZ. Through a combination of dissolution, diffusion, and buoyancy effects, methane transports into the GHSZ where the gas hydrate layer re-forms.
The hydrate layer follows the base of the GHSZ, but shrinks along the way as more and more gas dissolves and diffuses away.
In Fig.\ref{fig:hires_p2_results}, this process of hydrate dissociation $\rightarrow$ gas migration $\rightarrow$ hydrate reformation is shown in greater detail by zooming in on the time axis between $60,000$ years $\leq t \leq$ $120,000$ years.
These processes are quite complex and nonlinear. 
As the gas hydrate dissociates from the bottom, the free gas rises upwards with a decreasing speed due to a decreasing permeability in the hydrate phase.
When the gas phase crosses the region with maximum $S_h$, the speed of gas migration starts to increase until it escapes the hydrate layer on the other side, where a new hydrate layer starts to form, as shown in Fig.\ref{fig:hires_p2_Sh_Sg_t=0275}.
This new hydrate layer continues to grow using the free gas supplied by the dissociation of the old hydrate layer, as shown in Figs.\ref{fig:hires_p2_results}c,d,e.

\paragraph{}
In Fig.\ref{fig:p2_problemtime_vs_dt}, the evolution of $dt$ is plotted over the problem time for both schemes.
We can see that at $t=22,500$ years, when the gas phase first appears, $dt$ breaks down for both the schemes. However, the reduction of $dt$ for NCP scheme is not as severe as that for PVS scheme. 
The time step size gradually recovers as the gas slowly migrates upwards through the hydrate layer, but breaks down again around $t=60,000$ years, when the free gas crosses the region with peak $S_h$.

\begin{table}[tbh]
 \caption{Initial and bounday conditions for Example 2 (Sec. \ref{subsec:example2}).}
 \label{table:p2_ICBCs}
 \centering
    \begin{tabular}{| l !{\color{lightgray}\vrule} R{0.5cm} C{0.25cm} L{4cm} |}
    \hline
    \multicolumn{4}{|c|}{Initial conditions} \\
    \hline
    \multirow{16}{*}{at $t=0$, and $0$m $\geq z \geq$ $-800$m} 
    & $P_w $            & $ = $ & $P_{sf} + \rho_w g \left( z_{sf} - z\right)$           
    \\
    & \multicolumn{3}{L{6cm}|}{where, $z_{sf}=0$m is the sea floor, }
    \\
    & \multicolumn{3}{L{6cm}|}{and $P_{sf}=15$ MPa is the water pressure at the sea floor. }
    \\ [5ex]
    & $T $              & $ = $ & $ T_{sf} + d_z T_G \left( z_{sf}-z \right) $  
    \\
    & \multicolumn{3}{L{6cm}|}{where, $T_{sf}=4^0$C is the bottom water temperature,}\\
    & \multicolumn{3}{L{6cm}|}{and, $d_z T_G=35^0\text{C/km}$ denotes the regional geothermal temperature gradient. }
    \\ [4ex]
    & $S_g $            & $ = $ & $ 0 $                     
    \\ [1ex]
    & $\chi_w^{CH_4} $  & $ = $ & $ 0 $                     
    \\ [1ex]
    & $\chi_g^{H_2O} $  & $ = $ & $ \chi_{g,sat}^{H_2O}\left( \left.P_g\right|_{t=0}, \left.T\right|_{t=0} \right) $   
    \\ [1ex]
    & $\chi_w^c $       & $ = $ & $ 5.5 \text{ mmol/mol} $   
    \\ [1ex]
    \arrayrulecolor{lightgray}\hline\arrayrulecolor{black}
    at $t=0$,
    & & & \\
    \qquad if, \quad  $-320$m $\geq z \geq$ $-400$m ,
    & $S_h $           & $ = $ & $0.3 \left( \dfrac{400+z}{400-320} \right) \left( \dfrac{z+320}{400-320} \right)$ \\
    else if,   \quad  $z>-320$m or $z<-400$m
    &$S_h $            & $ = $ & $0$\\
    \hline
    \multicolumn{4}{|c|}{Boundary conditions} \\ 
    \hline
    \multirow{4}{*}{$t>0$, and $z = 0$m}
    & $P_w$         & $=$ & $P_{sf} + \rho_s \ g \ v_{s,z} \left(t_{n}+\Delta t \right)$ \\
    & $T$           & $=$ & $T_{sf} + d_z T_G \ v_{s,z} \left(t_{n}+\Delta t \right)$ \\
    & $S_g$         & $=$ & $0$ \\
    & $\chi_w^c$    & $=$ & $\left.\chi_w^c\right|_{t=0}$ 
    \\
    \arrayrulecolor{lightgray}\hline\arrayrulecolor{black}
    \multirow{4}{*}{$t>0$, and $z = -800$m}
    & $\partial_z P_w$        & $=$ & $0$ \\
    & $\partial_z T$          & $=$ & $d_z T_G$ \\
    & $v_{g,z}$               & $=$ & $0$ \\
    & $\partial_z\chi_w^c$    & $=$ & $0$ 
    \\
    \hline
    \end{tabular}
\end{table}

\subsection{Example 3: Gas production through depressurization}
\label{subsec:example3}
\paragraph{}
In this example, we numerically simulate a gas production scenario where the gas hydrate reservoir is destabilized through depressurization.
We consider a single-well configuration, and base the model parameters, material properties, and initial conditions within the reservoir on the geological setting of the Black Sea, similar to Example 2 (Sec.\ref{subsec:example2}).
The objective of this example is to demonstrate the robustness of our scheme in handling complex phase transitions even in those settings where the reaction rates are large, the hydrate phase distributions are highly heterogeneous, and the permeability typically varies over two to four orders of magnitude. 
Such settings are commonly found in natural gas hydrate systems which occur in turbidite formations containing thin hydrate layers sandwiched between thin layers of silty to clayey sediments.

\subsubsection{Problem setting}
\label{subsubsec:example3_problem_setting}

\begin{figure}[tbh]
 \centering
 \includegraphics[scale=0.6]{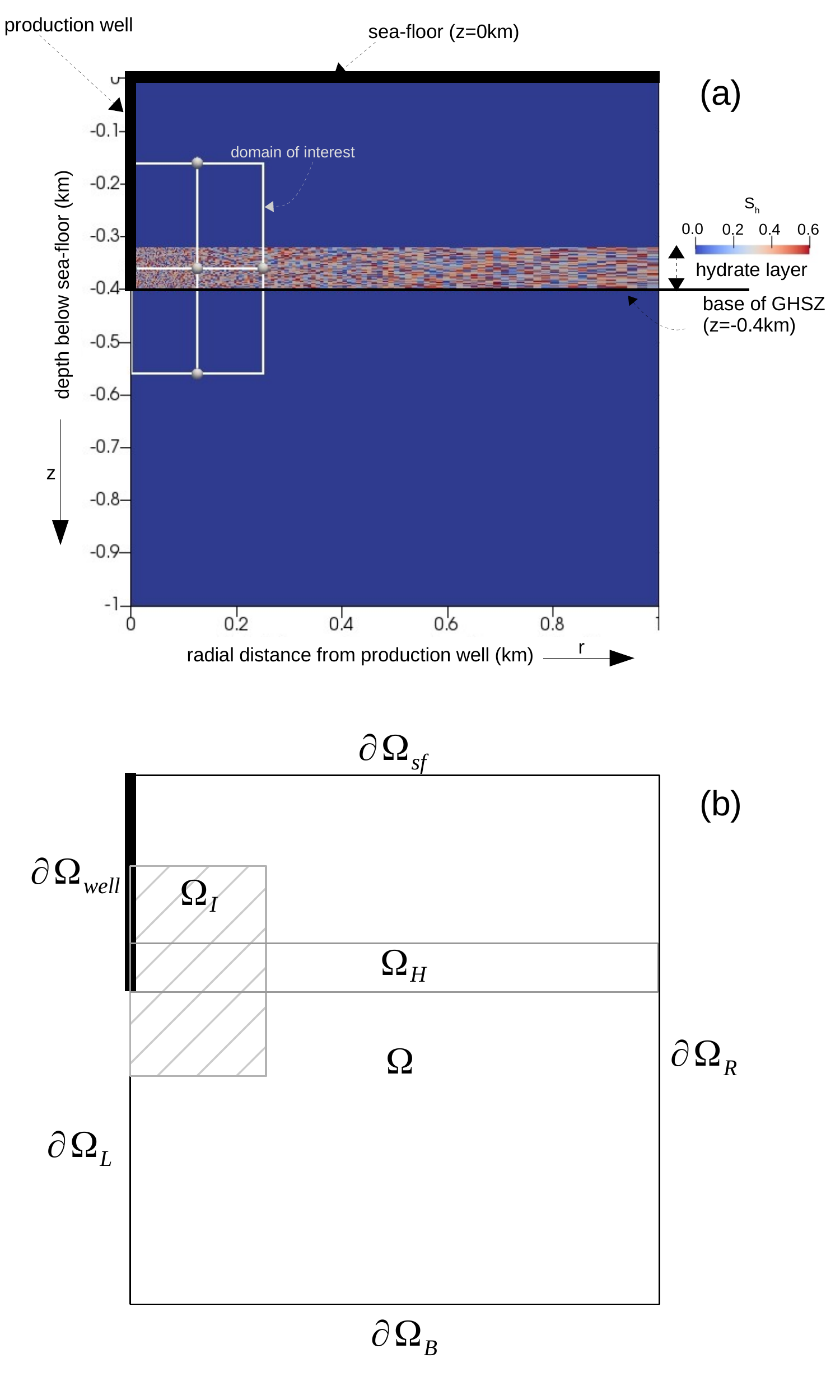}
 \caption{Problem setting for Example 3 (Sec.\ref{subsec:example3}).
          Figure (a) highlights the essential features of the problem setting like the locations of the sea floor, the production well, the initial base of the GHSZ and initial hydrate distribution within the hydrate layer, and marks our domain of interest within the computational domain.
          Figure (b) identifies the relevant regions of the computational domain.
          $\Omega$ denotes the computational domain, $\Omega_H \subset \Omega$ denotes the hydrate layer, $\Omega_I \subset \Omega$ denotes the domain of interest, and $\Omega_H\cap \Omega_I \neq \emptyset$.
          $\partial\Omega_{well}$ denotes the production well boundary,
          while $\partial\Omega_{L}$ denotes the left boundary excluding the production well.
          $\partial\Omega_{sf}$ denotes the top boundary corresponding to the sea floor. 
          $\partial\Omega_{R}$ and $\partial\Omega_{B}$ denote the right and the bottom boundaries, respectively.
%           Note that $\Omega_H$ and $\Omega_I$ are contained in $\Omega$, and $\Omega_H$ and $\Omega_I$ are overlapping.
%           Note, $\Omega_H \subset \Omega$ and $\Omega_I \subset \Omega$, and $\Omega_H\cap \Omega_I \neq \emptyset$.
         }
 \label{fig:p3_computational_domain}
\end{figure}

\paragraph{}
We consider a $2D$ axisymmetric domain, $\Omega$, having dimensions $1000$m $\times$ $1000$m, as shown in Fig. \ref{fig:p3_computational_domain}. 
The sea floor, $\partial\Omega_{sf}$, is prescribed at $z=0$m. 
% The depths below the sea floor are indicated with a $(-)$ sign.
The depressurizarion well, $\partial\Omega_{well}$, is located at $r=0$m, $0$m $\geq z \geq$ $-400$m.
The bottom water temperature at the sea floor is $T_{sf}=4^0$C, and the hydrostatic pressure at the sea floor is $P_{sf}=15$ MPa.
We assume that the absolute intrinsic permeability and the total porosity of the primary soil skeleton are $K=10^{-13} \text{ m}^2$ and $\phi=0.3$, respectively.

\paragraph{}
At $t=0$, we assume that the domain is fully saturated with saline water, and there is no free gas phase in the domain. Also, the aqueous phase contains no dissolved methane. 
For the aqueous phase pressure, we consider a hydrostatic pressure gradient, and for the temperature, we prescribe a regional geothermal gradient along the depth, $d_z T_G=35^0$C/km. 
The bGHSZ is located at $z=-400$m.
We consider that an $80$m thick gas hydrate layer, $\Omega_H$, exists right above the bGHSZ. 
To show the robustness of our scheme with respect to complex phase transitions, we prescribe a random distribution of the hydrate phase within this layer, s.t., the hydrate saturation ranges from $0$ to $0.6$, 
and the corresponding absolute permeability ranges from $ 10^{-13}\text{ m}^2$ to $ 1.6\times 10^{-15}  \text{ m}^2$.

\paragraph{}
For $t>0$, a pressure of $\left.P_w\right|_{\partial\Omega_{well}}=8$ MPa is prescribed at the production well to simulate gas production through depressurization.
At the sea floor, the temperature, pressure, and salinity conditions remain constant and equal to the initial values. 
At the bottom boundary, $\partial\Omega_B$, the regional geothermal gradient is maintained.

\paragraph{}
The initial and the boundary conditions are listed in Table \ref{table:p3_ICBCs},
and the material properties and model parameters are listed in Table \ref{table:material_properties_and_model_parameters}.

\subsubsection{Numerical simulation and results}
\label{subsubsec:example3_numerical_simulation_and_results}

\paragraph{}
The computational domain was discretized into a total of $20,276$ quadrilateral elements.
The gas production process aws simulated until $t_{end}=360$ days.
The maximum time step size was chosen as $dt_{max}=36,000$ sec,
and the time step size $dt$ was controlled adaptively using the heuristic strategy discussed in Sec. \ref{subsubsec:p1_numerical_simulation_and_results} with $\ell_{l}=4$ and $\ell_{h}=8$.
The simulation was run in parallel on $4$ processing units and required a total of $20$ CPU-hours.

\paragraph{}
We identify a domain of interest, $\Omega_I$, for the gas production process as: $0$m $\leq r \leq$ $250$m, $-150$m $\geq z \geq$ $-550$m.
Outside this domain, pressure, temperature, and saturation profiles do not change much.
The large size of the domain, however, is necessary to ensure that effects of depressurization do not reach $\partial\Omega_R$, and the geothermal gradient is maintained at $\partial\Omega_B$.

\begin{figure}[tbh]
    \subfigure[b]{
        \includegraphics[width=\textwidth,trim={0.2cm -0.75cm 0 7cm},clip]{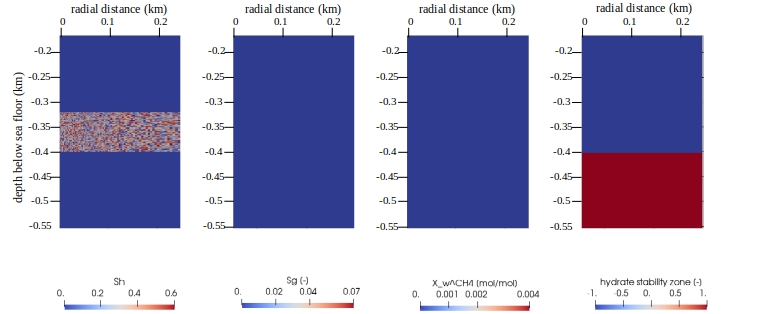}
    }
    \vspace{0.5cm}
    \subfigure[$t=10$ days]{
        \includegraphics[width=\textwidth,trim={0 2cm 0 0},clip]{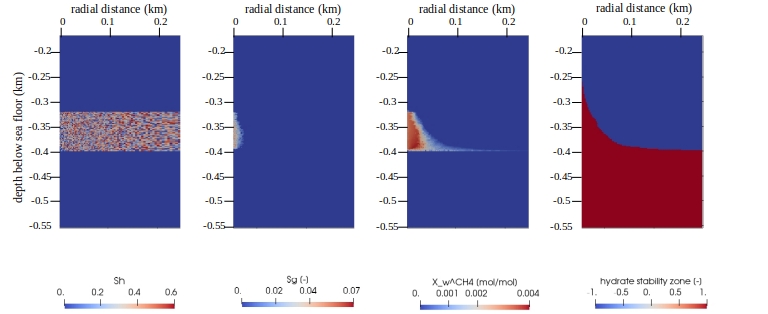}
        \label{fig:p3_Sh_Sg_XCH4_t=010}
    }
    \vspace{0.5cm}
    \subfigure[$t=90$ days]{
        \includegraphics[width=\textwidth,trim={0 2cm 0 0},clip]{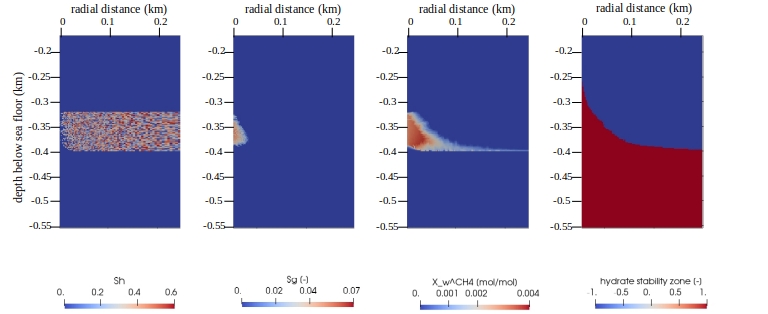}
        \label{fig:p3_Sh_Sg_XCH4_t=090}
    }
    \vspace{0.5cm}
    \subfigure[$t=360$ days]{
        \includegraphics[width=\textwidth,trim={0 2cm 0 0},clip]{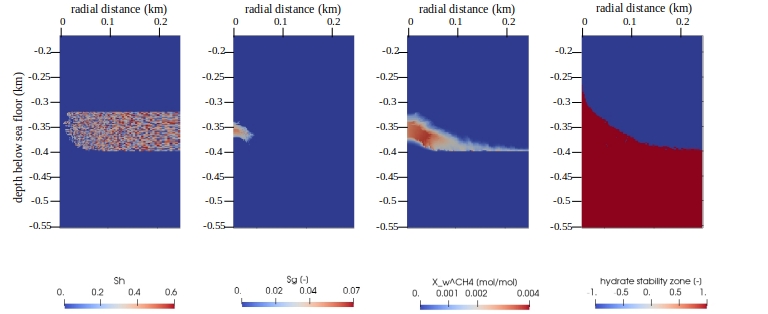}
        \label{fig:p3_Sh_Sg_XCH4_t=360}
    }
    \caption{Numerical results for Example 3 (Sec. \ref{subsec:example3}).
             The figure shows snapshots of the QoIs (from left to right: $S_h$, $S_g$, $\chi_w^{CH_4}$, and GHSZ) within the domain of interest $\Omega_I$ at different times. 
             Note, for GHSZ, a value of $1$ indicates unstable zone, and $-1$ indicates stable zone.}
    \label{fig:p3_results}
\end{figure}

\paragraph{}
The main quantities of interest (QoI) for gas production in gas hydrate reservoirs are $S_h$, $S_g$, $\chi_w^{CH_4}$, and the GHSZ.
The snapshots of the QoI within $\Omega_I$ are plotted in Fig.\ref{fig:p3_results} for times $t=10$ days, $t=90$ days, and $t=360$ days.
We can see that over a period of one year, roughly $100$ m of the reservoir is effectively depressurized, and the hydrate layer is fully dissociated within a zone of roughly $15$m around the production well. 
Due to the relatively high pore pressures, most of the methane is produced from the aqueous phase, and the saturation of the free gas phase remains well below $10\%$.

\begin{figure}[tbh]
\centering
    \includegraphics[width=\textwidth]{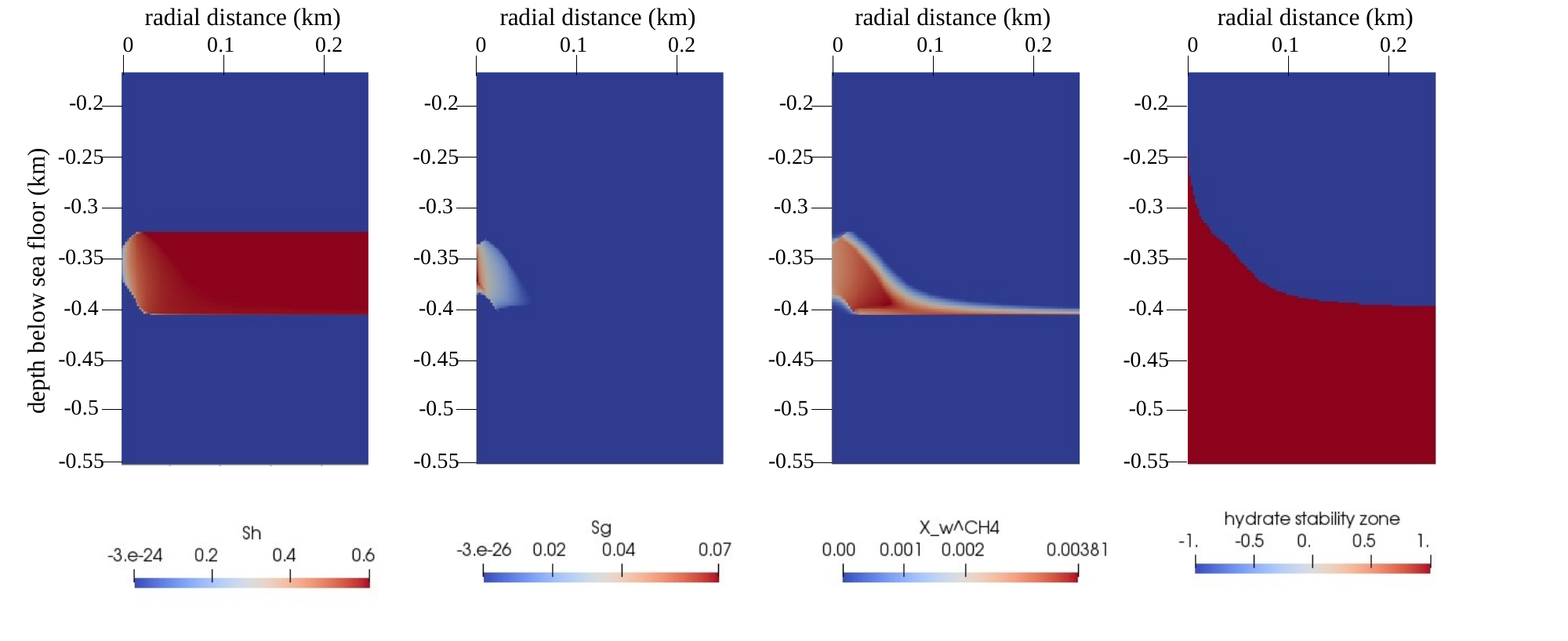}
    \caption{Numerical results for the reference test case of Example 3 (Sec. \ref{subsec:example3}).
             The figure shows snapshots of the QoIs (from left to right: $S_h$, $S_g$, $\chi_w^{CH_4}$, and GHSZ) within the domain of interest $\Omega_I$ at $t=t_{end}=360$ days. 
             Note, for GHSZ, $1$ indicates unstable zone, and $-1$ indicates stable zone.}
    \label{fig:p3_results_reference_case}
\end{figure}

\begin{figure}[tbh]
 \centering
 \includegraphics[scale=0.4]{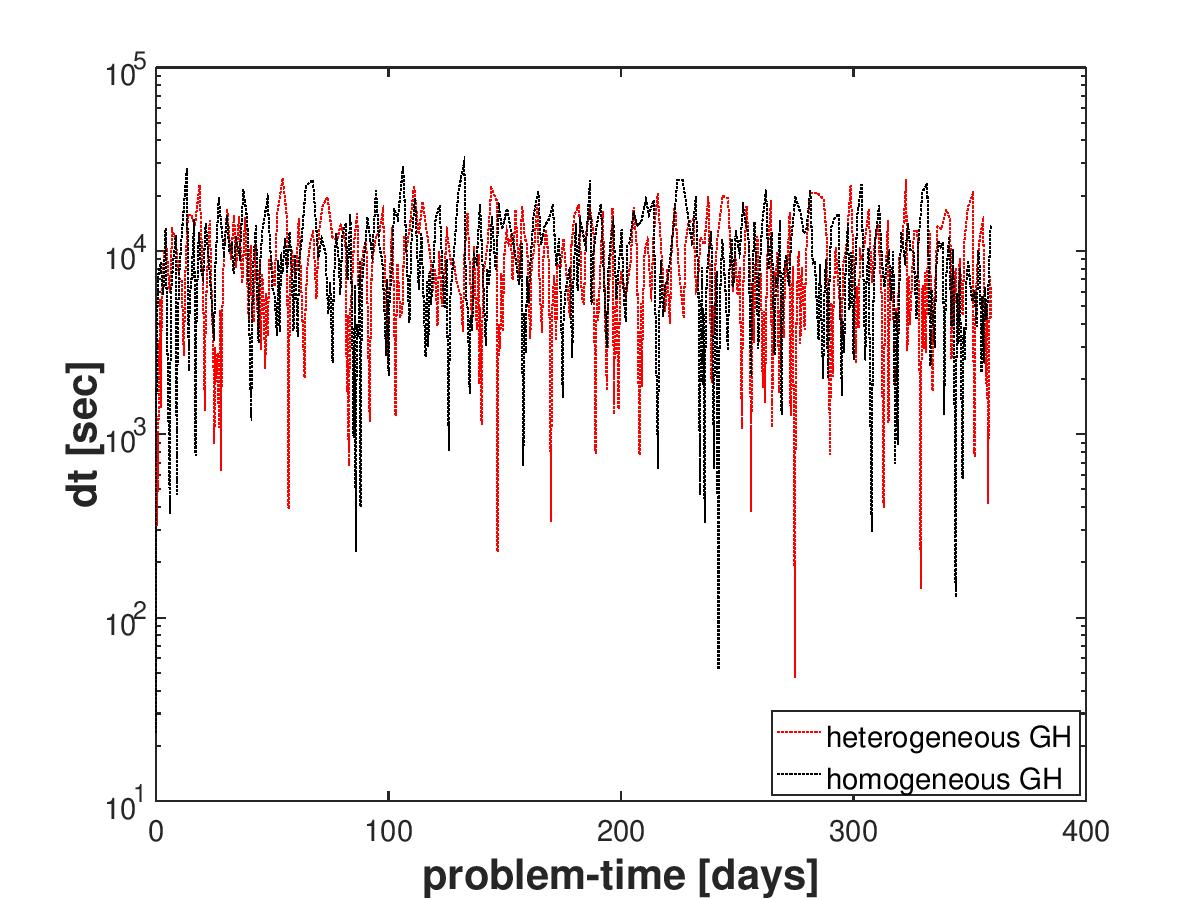}
 \caption{Numerical result for Example 3 (Sec. \ref{subsec:example3}).
 A comparison of the time step size evolution for a random hydrate phase distribution and a homogeneous hydrate phase distribution.}
 \label{fig:p3_dt_vs_problemtime}
\end{figure}

\paragraph{}
We compared the performance of the numerical scheme for this example with that of a \textit{reference} gas production test case.
The setting of the reference test case is the same as described in Sec.\ref{subsubsec:example3_problem_setting}, except that the hydrate distribution in $\Omega_H$ is homogeneous and has a uniform saturation of $0.6$.
A snapshop of the QoI in $\Omega_H$ for the reference test case is plotted in Fig.\ref{fig:p3_results_reference_case} at time step $t=360$ days. 
By comparing the behaviour of the numerical solution with the reference test case, we can ensure that the random hydrate distribution has not introduced any artificial numerical artifacts in the numerical solution.
In Fig.\ref{fig:p3_dt_vs_problemtime}, we can also see that despite the random distribution of the hydrate phase and the large variations in the sediment permeability, the semi-smooth Newton scheme is able to handle the phase transitions quite robustly without significant loss in performance, as indicated by the evolution of the time step sizes.

% {\setstretch{1.0}
\begin{table}[tbh]
 \caption{Initial and bounday conditions for Example 3 (Sec. \ref{subsec:example3}).}
 \label{table:p3_ICBCs}
 \centering
    \begin{tabular}{| L{6cm} !{\color{lightgray}\vrule} R{0.5cm} C{0.25cm} L{3cm} |}
    \hline
    \multicolumn{4}{|C{9.75cm}|}{Initial conditions} \\
    \hline
    \multirow{10}{*}{tag: $\Omega$} 
    & $P_w $            & $ = $ & $P_{sf} + \rho_w g \left( z_{sf} - z\right)$           \\
    \multirow{10}{*}{at $t=0$ ,} 
    & \multicolumn{3}{L{5.0cm}|}{where, $z_{sf}$ denotes the sea floor, $z_{sf}=0$m, }\\
    \multirow{10}{*}{$0$m $\leq r \leq$ $1000$m and $0$m $\geq z \geq$ $-1000$m}
    & \multicolumn{3}{L{5.0cm}|}{and $P_{sf}$ denotes the water pressure at the sea floor, $P_{sf}=15$ MPa. }\\ [5ex]
    & $T $              & $ = $ & $ T_{sf} + d_z T_G \left( z_{sf}-z \right) $  \\
    & \multicolumn{3}{L{5.0cm}|}{where, $T_{sf}=4^0$C denotes the temperature at the sea floor, }\\
    & \multicolumn{3}{L{5.0cm}|}{and, $d_z T_G=35^0\text{C/km}$ denotes the regional geothermal temperature gradient. }\\ [4ex]
    & $S_g $            & $ = $ & $ 0 $                     \\ [1ex]
    & $\chi_w^{CH_4} $  & $ = $ & $ 0 $                     \\ [1ex]
    & $\chi_g^{H_2O} $  & $ = $ & $ \chi_{g,sat}^{H_2O}\left( \left.P_g\right|_{t=0}, \left.T\right|_{t=0} \right) $   \\ [1ex]
    & $\chi_w^c $       & $ = $ & $ 5.5 \text{ mmol/mol} $   \\ [1ex]
    \arrayrulecolor{lightgray}\hline\arrayrulecolor{black}
    at $t=0$ , and  $0$m $\leq r \leq$ $1000$m ,
    & & & \\
    tag: $\Omega_H$: \quad\quad  $-320$m $\geq z \geq$ $-400$m
    & $S_h $           & $ = $ & rand$\left[ 0,0.6 \right]$ \\
    tag: $\Omega - \Omega_H$: \quad  $z>-320$m or $z<-400$m
    &$S_h $            & $ = $ & $0$\\
    \hline
    \multicolumn{4}{|C{9.75cm}|}{Boundary conditions} \\ 
    \hline
    tag: $\partial\Omega_{well}$
    & $P_w$ & $=$ & $8\text{ MPa}$ \\
    $t>0$, \quad $r=0$ and $0$m $\geq z \geq$ $-400$m
    & $\nabla T$ & $=$ & $0$ \\
    & $\mathbf{v}_g$ & $=$ & $0$ \\
    & $\nabla \chi_w^c$ & $=$ & $0$ \\
    \arrayrulecolor{lightgray}\hline\arrayrulecolor{black}
    tag: $\partial\Omega_{sf}$
    & $P_w$ & $=$ & $P_{sf}$ \\
    $t>0$, \quad $z=0$ and $0$m $\leq r \leq$ $1000$m
    & $T$ & $=$ & $T_{sf}$ \\
    & $S_g$ & $=$ & $0$ \\
    & $\chi_w^c$ & $=$ & $\left.\chi_w^c\right|_{t=0}$ \\
    \arrayrulecolor{lightgray}\hline\arrayrulecolor{black}
    tag: $\partial\Omega_{R}$
    & $\mathbf{v}_w$ & $=$ & $0$ \\
    $t>0$, \quad $r=1000$m and $0$m $\geq z \geq$ $-1000$m
    & $\nabla T$ & $=$ & $0$ \\
    & $\mathbf{v}_g$ & $=$ & $0$ \\
    & $\nabla \chi_w^cg$ & $=$ & $0$ \\
    \arrayrulecolor{lightgray}\hline\arrayrulecolor{black}
    tag: $\partial\Omega_{B}$
    & $\mathbf{v}_w$ & $=$ & $0$ \\
    $t>0$, \quad $z=-1000$m and $0$m $\leq r \leq$ $1000$m
    & $\nabla T$ & $=$ & $\nabla T_G$ \\
    & $\mathbf{v}_g$ & $=$ & $0$ \\
    & $\nabla \chi_w^cg$ & $=$ & $0$ \\
    \arrayrulecolor{lightgray}\hline\arrayrulecolor{black}
    tag: $\partial\Omega_{L}$
    & $\mathbf{v}_w$ & $=$ & $0$ \\
    $t>0$, \quad $r=0$m and $-400$m $> z \geq$ $-1000$m
    & $\nabla T$ & $=$ & $0$ \\
    & $\mathbf{v}_g$ & $=$ & $0$ \\
    & $\nabla \chi_w^cg$ & $=$ & $0$ \\
    \hline
    \end{tabular}
\end{table}
% }

%  trim={<left> <lower> <right> <upper>}

\begin{landscape}
\begin{table}
    \caption{Material properties and model parameters for Examples 1, 2, and 3.}
    \label{table:material_properties_and_model_parameters}
    \begin{footnotesize}
    \begin{center}
    \begin{tabular}{| L{1.4cm} !{\color{lightgray}\vrule} c !{\color{lightgray}\vrule} c !{\color{lightgray}\vrule} C{1.25cm} !{\color{lightgray}\vrule} C{1.25cm} !{\color{lightgray}\vrule} c |}
        \hline
        \multicolumn{3}{|c|}{Property$^{\text{2 3}}$}     
        & Example 1     & Example 2     & Example 3     \\
        \hline
        \multicolumn{6}{|c|}{\cellcolor{blue!5}{water}} \\
        \hline
        density                             & $\rho_w$          & $kg/m^3$  
        &   $1026.77$
        &   $1030.21$
        & $1027 + 0.45 P_w \left[MPa\right] - 0.15 \left(T\left[^0C\right]-10 \right) + 0.3521\times10^3 \left( \chi_w^c - 0.0096\right)$ 
        \\
        \arrayrulecolor{lightgray}\hline\arrayrulecolor{black}
        dynamic viscosity                   & $\mu_w$           & $Pa.s$    
        &   $0.0014$
        &   $0.00136$
        & $0.001792 \ \exp \left(\phantom{\left(\dfrac{}{}\right)^0} \hspace{-0.65cm} -1.94 \right. - \left. 4.80 \left(\dfrac{273.15}{T}\right) \right. + \left. 6.75 \left(\dfrac{273.15}{T}\right)^2\right)$
        \\
        \arrayrulecolor{lightgray}\hline\arrayrulecolor{black}
        thermal conductivity                & $k^{th}_w$        & $W/m/K$   
        &   $0.5854$
        &   $0.59$
        & $ 0.57153 \left( 1 \right. + \left. 0.003 T\left[^oC\right] \right. - \left. 1.025 \times 10^{-5} \left(T\left[^oC\right]\right)^2 \right. + \left. 6.53\times10^{-10}P_w \right. - \left. 0.0797\chi_w^c \right)$ 
        \\
        \arrayrulecolor{lightgray}\hline\arrayrulecolor{black}
        specific heat capacity              & $Cp_w$            & $J/kg/K$  
        & \multicolumn{3}{c}{$3945$}
        \\
        \arrayrulecolor{lightgray}\hline\arrayrulecolor{black}
        saturation vapour pressure          & $P_{sat}^{H_2O}$  & $Pa$      
        & $1072.92$  
        &\multicolumn{2}{c|}{$Pc \ \exp\left( \dfrac{1}{Tr} \left( c_1 (1-Tr) \right. \right. + \left. \left. c_2 (1-Tr)^{1.5} \right. \right. + \left. \left. c_3(1-Tr)^3 \right. \right. + \left. \left. c_4 (1-Tr)^{3.5} \right. \right. + \left. \left. c_5 (1-Tr)^4 \right. \right. + \left. \left. c_6 (1-Tr)^{7.5}\right) \phantom{\dfrac{}{}}\right)$} \\
        & & & &\multicolumn{2}{L{12cm}|}{ where, $Pc= 22.064$MPa, $Tc= 647.096$K, $Tr=T/Tc$, and $c_1=-7.85951783$, $c_2=1.84408259$, $c_3=-11.7866497$, $c_4=22.6807411$, $c_5=-15.9618719$, $c_6=1.80122502$. }
        \\
        \arrayrulecolor{lightgray}\hline\arrayrulecolor{black}
        diffusion coefficient               & $D^{H_2O}_g$      & $m^2/s$   
        &   $0.637\times 10^{-6}$
        &   $0.638\times 10^{-6}$
        & $2.26\times 10^{-9} T + \dfrac{0.002554}{P_g}$ 
        \\
        \hline
        \multicolumn{6}{|c|}{\cellcolor{blue!5}{methane}} \\
        \hline
        density                             & $\rho_g$          & $kg/m^3$  
        & $19.605$  
        & $0.002756\dfrac{P_g}{T}$  
        & $\dfrac{P_g}{z_{CH_4} R_{CH_4} T}$ \quad where, $R^{CH_4}=\frac{8314.5}{16.04}$, and $z^{CH_4}$ is estimated using Peng Robinson EoS \cite{PengRobinson1970}.  
        \\
        \arrayrulecolor{lightgray}\hline\arrayrulecolor{black}
       dynamic viscosity                   & $\mu_g$           & $Pa.s$    
        & $1.1045\times10^{-5}$  
        & $1.4055\times10^{-5}$  
        & $\mu_0 \left(\dfrac{273.15+162}{T+162}\right) \left( \dfrac{T}{273.15}\right)^{1.5}$ \\
        & & & & & where, $\mu_0 = 1.0707\times10^{-5} - 4.8134\times10^{-14}P_g - 4.1719\times10^{-20} P_g^2 + 7.3232\times10^{-28} P_g^3 $
        \\
        \arrayrulecolor{lightgray}\hline\arrayrulecolor{black}
        thermal conductivity                & $k^{th}_g$        & $W/m/K$   
        & $0.03107$  
        & $0.03121$  
        & $ a_0 + a_1 T + a_2 T^2 + a_3 T^3 $ \\
        & & & & & where, $a_0=-0.008863$, $a_1=0.000242$, $a_2=-0.6997\times10^{-6}$, and $a_3=0.1225\times10^{-8}$.
        \\
        \arrayrulecolor{lightgray}\hline\arrayrulecolor{black}
        specific heat capacity              & $Cp_g$            & $J/kg/K$  
        & $2165.24$ 
        & $2168.65$  
        & $ 1238 + 3.13 T + 7.905\times10^{-4} T^2 - 6.858\times10^{-7} T^3$ 
        \\
        \arrayrulecolor{lightgray}\hline\arrayrulecolor{black}
        solubility constant                 & $H_w^{CH_4}$      & $Pa$      
        & $1,343\times 10^{11}$  
        & \multicolumn{2}{c|}{$\exp \left( \log(P_{sat}^{H_2O}) - \dfrac{11.0094}{Tr} + 4.8362 \dfrac{(1-Tr)^{0.355}}{Tr} + 12.5220 \ \exp(1-Tr) Tr^{-0.41} \right)$} 
        \\
        \arrayrulecolor{lightgray}\hline\arrayrulecolor{black}
        diffusion coefficient               & $D^{CH_4}_w$      & $m^2/s$   
        & $1.57\times10^{-11}$
        & $1.57\times10^{-11}$  
        & $1.57 \times 10^{-11} \left(\dfrac{P_w}{1.0135 \times 10^{5}}\right) \exp\left({-\dfrac{0.003475}{T}}\right) $ \\
        \hline
    \end{tabular}
    \end{center}
    \end{footnotesize}
\end{table}
\end{landscape}

\begin{table}
    \centering
    \caption*{(Table \ref{table:material_properties_and_model_parameters} continued.) Material properties and model parameters for Examples 1, 2, and 3. }
    \begin{footnotesize}
    \begin{tabular}{| L{1.5cm} !{\color{lightgray}\vrule} c !{\color{lightgray}\vrule} c !{\color{lightgray}\vrule} C{1.25cm} !{\color{lightgray}\vrule} C{1.25cm} !{\color{lightgray}\vrule} C{3cm} |}
        \hline
        \multicolumn{3}{|c|}{Property
        \footnote{See \cite{Kossel2013} for references to $P^{H_2O}_{sat}$ and $H_w^{CH_4}$, 
                  \cite{Janicki2011} for references to $k^{th}_w$, $Cp_w$, $\mu_{0,g}$, $\rho_h$, $N_h$, $k^{th}_h$, $Cp_h$, $D^c_w$, $\rho_s$, $k^{th}_s$, and $Cp_s$,
                  and \cite{Gupta2017} for references to $\mu_w$, $D_g^{H_2O}$, $\mu_g$, $D_w^{CH_4}$, $k^{th}_g$, $Cp_g$, and ${\dot Q}_h$.
                  }
                  \footnote{To evaluate the property values for Example 1, we considered a reference state of $T=8^0$C, $P=2$ MPa and $\chi_w^c=5.5$ mmmol/mol,
                  and for Example 2, we considered a reference state of $T=9^0$C, $P=10$ MPa and $\chi_w^c=5.5$ mmmol/mol.
                  }
        }   
        & Example 1     & Example 2      & Example 3     \\
        \hline
        \multicolumn{6}{|c|}{\cellcolor{blue!5}{hydrate}} \\
        \hline
        density                             & $\rho_h$          & $kg/m^3$  
        & \multicolumn{3}{c|}{$920$} 
        \\
        \arrayrulecolor{lightgray}\hline\arrayrulecolor{black}
        hydration number                    & $N_h$             & $-$       
        & \multicolumn{3}{c|}{$5.90$} 
        \\
        \arrayrulecolor{lightgray}\hline\arrayrulecolor{black}
        thermal conductivity                & $k^{th}_h$        & $W/m/K$   
        & \multicolumn{3}{c|}{$0.5$} 
        \\
        \arrayrulecolor{lightgray}\hline\arrayrulecolor{black}
        specific heat capacity              & $Cp_h$            & $J/kg/K$  
        & $2216$
        & $2327$ 
        & $\left(1.937 T^3 \right. - \left. 1.5151 T^2 \right. + \left. 3.9554 T \right. - \left. 342.7\right)\times10^3 $\\
        \hline
        \multicolumn{6}{|c|}{\cellcolor{blue!5}{salt}} \\
        \hline
        diffusion coefficient               & $D^{c}_w$         & $m^2/s$   
        & \multicolumn{3}{c|}{$10^{-9}$} 
        \\
        \hline
        \multicolumn{6}{|c|}{\cellcolor{blue!5}{soil}} \\
        \hline
        density                             & $\rho_s$          & $kg/m^3$  
        & \multicolumn{3}{c|}{$2600$} 
        \\
        \arrayrulecolor{lightgray}\hline\arrayrulecolor{black}
        thermal conductivity                & $k^{th}_s$        & $W/m/K$   
        & \multicolumn{3}{c|}{$3.0$}
        \\
        \arrayrulecolor{lightgray}\hline\arrayrulecolor{black}
        specific heat capacity              & $Cp_s$            & $J/kg/K$  
        & \multicolumn{3}{c|}{$1000$} 
        \\
        \hline
        \multicolumn{6}{|c|}{\cellcolor{blue!5}{hydrate phase change kinetics}} \\
        \hline
        hydrate equilibrium pressure        &  $P_{e}$          & $Pa$      
        & \multicolumn{3}{c|}{$\exp\left( 38.592 - \dfrac{8533.8}{T} + 4.4824 \chi_w^c \right)$}  
        \\
        \arrayrulecolor{lightgray}\hline\arrayrulecolor{black}
        kinetic rate constant  & $k^r $  & $\text{mol}/m^2/Pa/s$  
        & $10^{-12}$ 
        & $10^{-17}$  
        & $10^{-12}$ 
        \\
        \arrayrulecolor{lightgray}\hline\arrayrulecolor{black}
        specific surface area               & $A_0$         & $m^2/m^3$     
        & \multicolumn{3}{c|}{$10^5$}  
        \\
        \arrayrulecolor{lightgray}\hline\arrayrulecolor{black}
        heat of reaction                    & ${\dot Q}_h$  & $W/m^3$       
        &  \multicolumn{3}{c|}{$\dfrac{\dot g_{h}}{M_h} \left( 56599-16.744 \ T \right)$}  
        \\
        \hline
        \multicolumn{6}{|c|}{\cellcolor{blue!5}{hydraulic properties}} \\
        \hline &&\\ [-2ex]
        absolute intrinsic permeability     & $K_0$                 & $m^2$     
        & $10^{-12}$    
        & $10^{-15}$ 
        & $10^{-13}$  
        \\
        \arrayrulecolor{lightgray}\hline\arrayrulecolor{black}
        total porosity                      & $\phi$                & $-$       
        & $0.3$         
        & $0.5$ 
        & $0.3$   
        \\
        \arrayrulecolor{lightgray}\hline\arrayrulecolor{black}
        Brooks-Corey parameters             & $p_0$, $\lambda$      & $Pa$,$-$ 
        & \multicolumn{3}{c|}{$5\times 10^4$, $1.2$}
        \\
        \arrayrulecolor{lightgray}\hline\arrayrulecolor{black}
        sphericity parameter                & $m$                   & $-$       
        & \multicolumn{3}{c|}{$1$} 
        \\
        \arrayrulecolor{lightgray}\hline\arrayrulecolor{black}
        residual saturations                & $S_{wr}$, $S_{gr}$    & $-$,$-$   
        & \multicolumn{3}{c|}{$0$,$0$}
        \\
        \hline
    \end{tabular}
    \end{footnotesize}
\end{table}

\section{Conclusion}
\label{sec:discussion_and_outlook}
\paragraph{}
In this article, we presented a mathematical model for non-isothermal multi-phase multi-compo\-nent reactive transport processes in methane hydrate reservoirs. 
The methane hydrate phase transitions were modelled as a non-equilibrium based kinetic process, and the phase transitions within the $CH_4-H_2O$ system were modelled as a VLE process.
The inequality conditions resulting from the VLE assumption were cast as KKT equality conditions which were implemented within a semi-smooth Newton scheme using an active-set strategy.
Note that, in the context of gas hydrate models, a similar nonlinear complementary constraints approach was also used by \cite{Gibson2014} to develop a semi-smooth Newton strategy for solving a Stefan-type problem involving equilibrium based hydrate phase transition. 

\paragraph{}
In Example 1 (Sec.\ref{subsec:example1}), we validated our semi-smooth Newton scheme against a PVS scheme by simulating a sequence of phase transitions involving appearance and disappearance of the gas phase over time.

\paragraph{}
In many widely used multi-phase multi-compo\-nent gas hydrate reservoir simulators, PVS is the method of choice for handling phase transitions and the vanishing gas phase.
In general, the PVS method has the advantage that the numerical model has fewer degrees of freedom, and therfore, can perfom numerical calculations faster. 
However, in the case of gas hydrate models, this advantage is most often lost because the phase transitions are highly coupled and the PVS scheme requires much smaller time-steps for convergence.
We demonstrated this in Example 2 (Sec.\ref{subsec:example2}) where we simulated gas migration through GHSZ in the highly dynamic geological setting of the Black Sea over paleo time-scales.
Notice in Table \ref{table:material_properties_and_model_parameters} that, we greatly simplified the problem setting for Example 2 by neglecting the functional dependence of the material properties on local temperature, pressure and salinity conditions, thereby, reducing the nonlinearities, and we considered only a $1D$ setting with homogeneous phase distributions across the domain. Despite these simplifications, the PVS scheme performed relatively poorly compared to the semi-smooth Newton scheme.
In Example 3 (Sec.\ref{subsec:example2}), we considered another very important application of methane hydrate models, viz., gas production through depressurization. 
In this example, we considered a random distribution of the hydrate phase and included strongly nonlinear functional dependencies of the material properties on the local thermodynamic state (see Table \ref{table:material_properties_and_model_parameters}) in order to show that our semi-smooth Newton scheme can robustly handle even very complex field-scale problems.

\begin{acknowledgements}
This project was funded by the Cluster of Excellence ``The Future Ocean''. The Future Ocean is funded within the framework of the Excellence Initiative by the Deutsche Forschungsgemeinschaft
(DFG) on behalf of the German federal and state governments.
\end{acknowledgements}

\bibliographystyle{spmpsci}      % mathematics and physical sciences
\bibliography{refs}   % name your BibTeX data base

% % Non-BibTeX users please use
% \begin{thebibliography}{}
% %
% % and use \bibitem to create references. Consult the Instructions
% % for authors for reference list style.
% % Format for Journal Reference
% Author, Article title, Journal, Volume, page numbers (year)
% Author, Book title, page numbers. Publisher, place (year)
% \end{thebibliography}

\end{document}